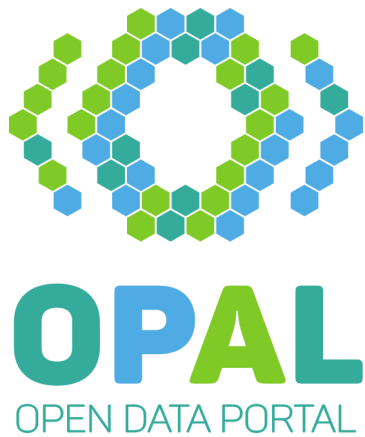

# Open Data Portal Germany (OPAL)
# Projektergebnisse

mFUND
*Das Startkapital für die Mobilität 4.0*

Bundesministerium
für Verkehr und
digitale Infrastruktur

# Inhaltsverzeichnis

# 1  Online Ressourcen

## 1.1  OPAL Projekt

- **Fachgruppe Data Science: OPAL**
  https://dice-research.org/OPAL
- **OPAL Projektwebseite**
  http://projekt-opal.de/
- **BMVI: OPAL**
  https://www.bmvi.de/SharedDocs/DE/Artikel/DG/mfund-projekte/ope-data-portal-germany-opal.html

## 1.2  Deliverables

- **GitHub**
  https://github.com/projekt-opal/doc/tree/master/deliverables
- **OPAL Projektwebseite (Backup)**
  http://projekt-opal.de/projektergebnisse/deliverables/
- **FTP Server Hobbitdata (Backup)**
  https://hobbitdata.informatik.uni-leipzig.de/OPAL/Deliverables/

## 1.3  Demo

- **Webseite zur Demo**
  https://dice-research.org/OPAL-Demo
- **OPAL Demo**
  https://opal.demos.dice-research.org/

## 1.4  Code

- **GitHub**
  https://github.com/projekt-opal/doc#repositories

## 1.5  Daten

- **FTP Server Hobbitdata**
  https://hobbitdata.informatik.uni-leipzig.de/OPAL/

# 2 Ergebnisse der Arbeitspakete

## 2.1 Arbeitspaket 1: Anforderungsanalyse und Architektur

**Ziel des Arbeitspaketes**

Ziel dieses Arbeitspakets ist die Erfassung der Anforderungen, die Untersuchung und initiale Analyse der Datenbestände der Entwurf einer Gesamtarchitektur für OPAL.

**Deliverables des Arbeitspaketes**

Die folgenden Abgaben wurden in diesem Arbeitspaket fertiggestellt:

- D1.1 Anforderungsanalyse
- D1.2 Datenanalyse
- D1.3 Architektur

### 2.1.1 Arbeitspaket 1.1: Erhebung der Nutzeranforderungen

**Anforderungsanalyse**

Neben einer systematischen Literaturrecherche wurden Anforderungen aus den Ergebnissen von Fragebögen und eines mFUND Workshops sowie Erkenntnissen aus der Fachkonzeption und einer Usability-Untersuchung des mCLOUD-Portals gewonnen. Das Ergebnis ist die folgende Liste konsolidierter Anforderungen.

**Konsolidierte Anforderungen (AK)**

1. Semantische Suche. Mit dieser Anforderung wird ein umfangreiche Suchfunktionalität über den verknüpften und einheitlichen Daten mit hoher Qualität ermöglicht.
2. Räumliche Suche. Die typischerweise räumlich begrenzten Daten aus mCLOUD/MDM und verwandten Portalen müssen für die Suchfunktionalität (AF1) geeignet gefiltert werden können.
3. Zeitliche Suche. Über das kontinuierliche Crawling und die extrahierten und in ein einheitliches Format transformierten Metadaten wird die Einschränkung auf Zeiträume ermöglicht.
4. Monitoring und Benachrichtigung. Das kontinuierliche Crawling ermöglicht die Überwachung der Verfügbarkeit der Daten sowie die Information von Nutzern über Zeitverlauf und Aktualisierungen.
5. Entwicklung eines komponentenbasierten Metadatenportals.
6. Programmatischer Zugriff über standardisierte API. Wenn möglich sollen mehrere Zugriffsmethoden und Datenformate bereitgestellt werden, um den unterschiedlichen Nutzeranforderungen zu genügen.
7. Konvertierung der Metadaten in Linked Data und unterschiedliche Formate und Vokabulare
8. Analyse der (Meta-)Datenqualität. Die Qualität der Daten wird mit geeigneten generischen und spezifischen Metriken untersucht.
9. Automatisierte Verknüpfung von Datensätzen.
10. Prüfung und Integration von Lizenzinformationen. Motivation ist die hohe Bedeutung korrekter Lizenzinformationen, als auch die Notwendigkeit, resultierende Lizenzbedingungen aus der Integration von mehreren Datensätzen abzuleiten.
11. Fokussierter Crawler, der relevante Webseiten mit offenen Daten anhand bereitgestellter Seed-Listen findet und untersucht
12. Automatisierte Extraktion von Metadaten aus relevanten Datenquellen anhand von Beispieldaten
13. Möglichkeit der Selektion von Teilmengen eines Datensatzes für ausgewählte Datenquellen
14. Mobile Anwendung zur Suche nach lokal relevante Datensätzen
15. Question-Answering-basierter Assistent für soziale Netzwerke
16. Untersuchung der Datensätze nach Datenportal, Datenanbieter, und weiteren Facetten, die mit logischen UND- und ODER-Operatoren kombiniert werden können
17. Persistente versionierte Speicherung von Metadaten
18. Anzeige von existierenden und neuen Daten im Repository, einschließlich verfügbarer Metadaten.
19. Empfehlungen von relevanten Datensätzen basierend auf semantischer Dokumentenanalyse, Suchanfrage, gewählten Datensätze und persönlichem Profil.

20. Kommentierung bezüglich Qualitäts- und inhaltsbezogener Kriterien.
21. Möglichkeit zur Bewertung der Datensätze anhand von nutzerzentrierten Qualitätskriterien.

**Weiterführende Inhalte**

- D1.1 Anforderungsanalyse (Matthias Wauer, Adrian Wilke): https://github.com/projekt-opal/doc /blob/master/deliverables/OPAL_D1.1_Anforderungsanalyse.pdf

### 2.1.2 Arbeitspaket 1.2: Datenanalyse

In der initialen Datenanalyse wurde eine Liste von relevanten Datenquellen identifiziert. Diese setzt sich aus den Metadatenportalen mCloud, MDM, GovData, OffeneDaten.de und dem European Data Portal zusammen. Die Datenquellen wurden nach multi-methodischem Ansatz analysiert.

Die Portale wurden mit unterschiedlichen Methoden betrachtet. Neben einer manuellen Prüfung bzw. einem Studium der Dokumentation der Datenquellen wurden umfangreichere Analysen auf dem Datenbestand der jeweiligen Quelle durchgeführt. Dies setze eine Erfassung der Meta-Datensätze voraus, durch vollständiges Crawling oder durch repräsentatives Random Sampling.

Als Ergebnis der Analyse lässt sich festhalten, dass das Crawling der Datenquellen (bei MDM eingeschränkt) möglich ist und eine Integration und Fusion der Metadaten aus den primären und externen Quellen durchführbar und sinnvoll erscheint.

**Anzahl der Datensätze in 2018**

Im Deliverable D1.2 zur Datenanalyse vom Mai 2018 wurden die folgenden Datensätze ermittelt:

| Datenquelle | Anzahl Datensätze |
|---|---|
| mCLOUD | 652 |
| MDM | 119 |
| GovData | 19.754 |
| OffeneDaten.de | 28.542 |
| European Data Portal | 817.755 (206.068 aus Deutschland) |

**Änderungen im weiteren Projektverlauf**

Während des Projektverlaufs hat sich einerseits die Anzahl der verfügbaren Datensätze erhöht. Andererseits ergaben sich technische Veränderungen. In 2,5 Jahren (Mai 2018 – Dezember 2020) wuchs der Datenbestand in der mCloud von 652 auf 3.276 an; beim European Data Portal von 817.000 auf 1.184.000. Technisch wird nach einem MDM Relaunch kein jQuery mehr genutzt und mCLOUD bietet seit Version 1.5.0 einen Download im DCAT-AP.de XML/RDF Format an, sodass sich das Scraping vereinfacht hat.

**OPAL Graph**

Der OPAL Graph in der Version von Oktober/November 2020 umfasst nach einer Datenbereinigung die folgende Anzahl Datensätze:

| Datenquelle | Anzahl Datensätze |
| --- | --- |
| MDM | 203 |
| mCLOUD | 2.853 |
| GovData | 37.932 |
| European Data Portal | 795.387 (191.374 mit deutschen und englischen Titeln) |

**Weiterführende Inhalte**

- D1.2 Datenanalyse (Matthias Wauer, Adrian Wilke): https://github.com/projekt-opal/doc/blob/master/deliverables/OPAL_D1.2_Datenanalyse.pdf
- Opal Graph: https://hobbitdata.informatik.uni-leipzig.de/OPAL/OpalGraph/

### 2.1.3 Arbeitspaket 1.3: Architektur

Eine High-Level Ansicht der OPAL-Architektur ist in der folgenden Abbildung dargestellt. Sie ist unterteilt in die folgenden Schichten:

1. Die Extraktionsschicht enthält die Funktionalität zum gezielten Crawlen von Open-Data-Seiten.
2. Die Datenanalyseschicht enthält ein erweiterbares Framework mit Komponenten zum Extrahieren bestimmter Metadatenelemente, die Verarbeitung von Qualitätsmetriken und die Extraktion von Schemainformationen Informationen für Datensätze.
3. Die Transformationsschicht überträgt die verschiedenen Metadaten-Repräsentationen in ein gemeinsames Vokabular.
4. Die Integrationsschicht verknüpft und vereinheitlicht verschiedene Metadatenbeschreibungen aus unterschiedlichen Portalen für einzelne Datensätze und berechnet Beziehungen zwischen einzelnen Datensätzen.
5. Die Zugriffsschicht bietet eine API für den programmatischen Zugriff auf die Portalfunktionalität.
6. Die Anwendungsschicht enthält die Endbenutzeranwendungen

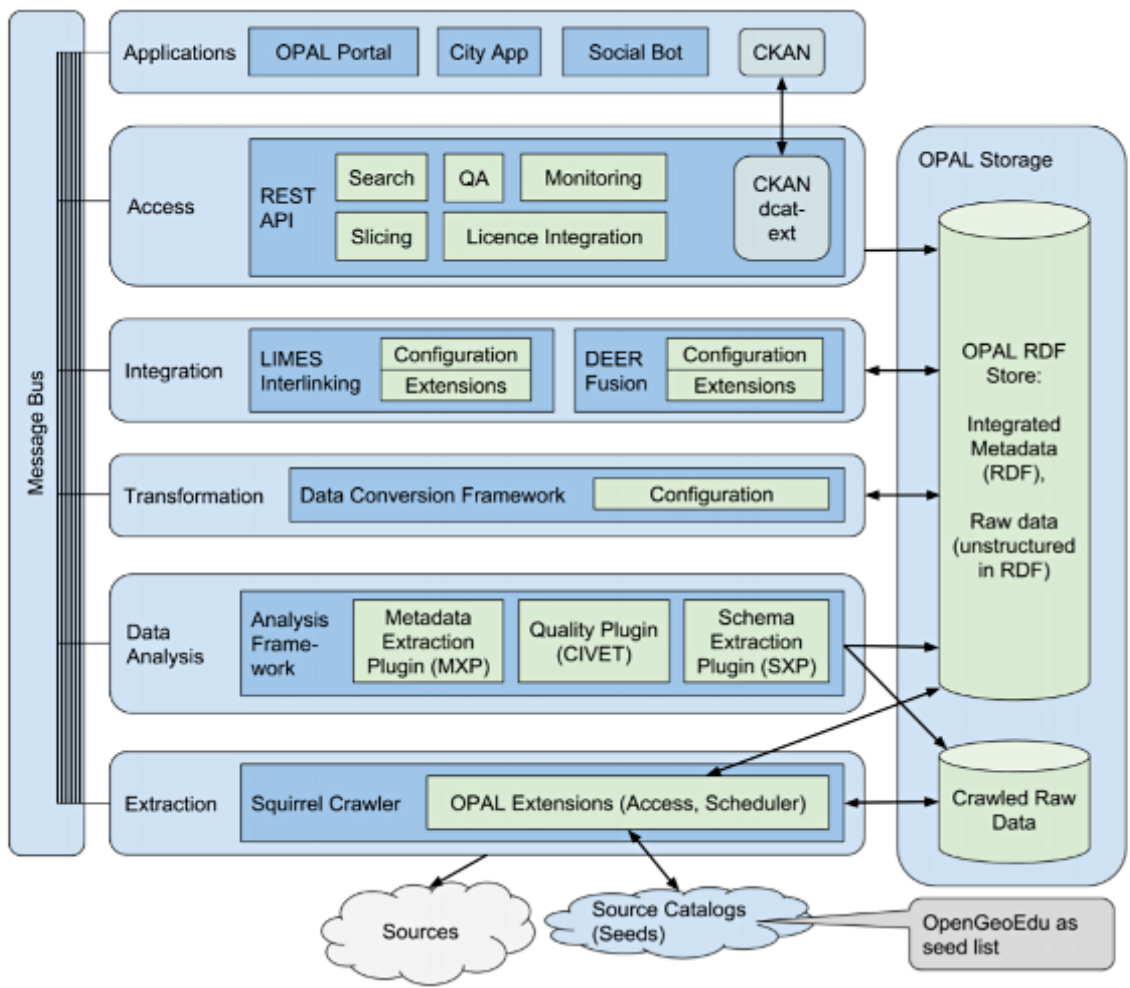

Während des Projektverlaufs hat sich die Verwendung der Datenhaltungsschicht als Flaschenhals herausgestellt. Daher wurde die ursprüngliche Architektur so umgestellt, dass für das Web-Frontend

Elasticsearch zum Einsatz kommt und zur Aufbereitung von Daten eine sequenzielle Behandlung erfolgt (siehe Arbeitspaket 4.2: Konvertierungskomponente).

**Weiterführende Inhalte**

- D1.3 Architektur (Matthias Wauer, Ivan Ermilov): https://github.com/projekt-opal/doc/blob/master/deliverables/OPAL_D1.3_Architecture.pdf

## 2.2  Arbeitspaket 2: Datenakquisition

**Ziel des Arbeitspaketes**

In Arbeitspaket 2 wird ein fokussierter Crawler entwickelt, der Informationen zu offenen Datensätzen aus Webseiten extrahiert.

**Deliverables des Arbeitspaketes**

Die folgenden Abgaben wurden in diesem Arbeitspaket fertiggestellt:

- D2.1 Spezifikation der Crawler-Komponente
- D2.2 Erste Version der Crawler-Komponente
- D2.3 Benchmark-Spezifikation und Ergebnisse des ersten Crawlers
- D2.4 Metadatenbasierte Crawlingstrategien
- D2.5 Finale Crawler-Komponente
- D2.6 Finale Crawler-Benchmark-Ergebnisse

### 2.2.1 Arbeitspaket 2.1: Spezifikation der Crawler-Komponente

Das primäre Ziel von OPAL ist es, Benutzern das einfache Auffinden offener Daten zu ermöglichen, indem verfügbare Metadaten aus verschiedenen Webquellen gesammelt, analysiert und integriert werden. Dieses Arbeitspaket konzentriert sich auf den ersten Schritt in Richtung dieses Ziels: Die Identifizierung und Extraktion von Metadaten von Webseiten, die Datensätze beschreiben.

Dieses Arbeitspaket spezifiziert funktionale und nicht-funktionale Anforderungen für die OPAL Crawler-Komponente. Dabei werden die allgemeinen Anforderungen, die in Deliverable D1.1 gesammelt wurden, referenziert. Außerdem werden 10 existierende Crawling-Komponenten hinsichtlich der Anforderungen verglichen.

**Funktionale Anforderungen**

Im Folgenden sind funktionale Anforderungen, wie Dienste oder Funktionen, die von der Crawler-Implementierung bereitgestellt werden müssen, aufgeführt. Zusätzlich sind die Anforderungen aus AP 1.1 referenziert.

| ID | Titel | Beschreibung | Bewertungskriterien | Ref. |
|----|-------|--------------|---------------------|------|
| CF1 | Fokussiertes Crawling von HTTP-Seiten | Der fokussierte Crawler verwendet vordefinierte Seed-Listen, um auf Web-Ressourcen zuzugreifen, diese zu durchlaufen und bietet Filtermöglichkeiten für die Verfolgung von Links zu anderen Web-Ressourcen. | Manueller Abgleich von Meta-Informationen bei Datenportalen mit den gecrawlten Daten. | AK11 |
| CF2 | Zugriff über verschiedene Protokolle | Der Crawler kann auf Daten über verschiedene Standardprotokolle zugreifen, z. B. HTTPS und FTP | Testfälle für Crawler-Konnektoren; Verfügbarkeit von gecrawlten Daten von Seiten, die diese Protokolle verwenden. | AK11 |
| CF3 | Periodisches Crawling | Funktion zum Starten von periodischen und iterativen Crawling-Prozessen, die vom Crawling-Framework unterstützt wird, ist in der OPAL-Administration verfügbar. | Verfügbarkeit dieser Funktion mit Unit-/Integrationstest. | AK4 |
| CF4 | Fetching | Relevante Rohdaten, die vom Crawler gefunden wurden, werden extrahiert und gespeichert. | Präzisions-/Recall-Evaluierung mit manuell annotierten Goldstandard-Stichproben von Quellen. | AK11 |
| CF5 | Analyse | Halbstrukturierte, geladene Daten werden analysiert, um nur auf relevante Teile zuzugreifen. | Präzisions-/Recall-Auswertung mit manuell annotierten Stichproben | AK12 |
| CF6 | Extraktion | Rohdatenstrings werden aus umgebenden Strukturdaten (z. B. HTML-Tags) extrahiert. | Extrahierte Daten enthalten keine Strukturinformationen. | AK12 |
| CF7 | Semantische, maschinenlesbare Datenspeicherung | Gefundene Daten sollten in einem Datenformat gespeichert werden, das geeignet ist, die Semantik der extrahierten Informationen zu handhaben, um sie in einem nachfolgenden Integrationsschritt zu verknüpfen. (z.B. Turtle, Terse RDF Triple Language) | Validierung der gecrawlten Daten. | AK7, AK9 |
| CF8 | Speicherung von zeitbezogenen Daten | Die Crawler-Komponente reichert extrahierte Daten (siehe CF7) mit einem Zeitstempel an und speichert sie. | Unit-Test. | AK17, AK4 |
| CF9 | CKAN-API-Unterstützung | Der Crawler kann die CKAN-API für einen effizienteren Datenzugriff nutzen. | Integrationstestfall. | - |

**Nicht-funktionale Anforderungen**

Nicht-funktionale Anforderungen beschreiben Qualitätsaspekte, Rahmenbedingungen oder Anforderungen, die auf verschiedene Weise umgesetzt werden können. In der folgenden Tabelle werden diese bzgl. eines fokussierten Crawlers beschrieben, wobei eine Referenzierung auf AP 1.1 erfolgt.

| ID | Titel | Beschreibung | Bewertungskriterien | Ref. |
|---|---|---|---|---|
| CN1 | Konfiguration | Eine Konfiguration von Daten-Seeds (z. B. Katalogübersichten) ist über eine Benutzeroberfläche für Menschen und auf programmatische Weise (über eine API) möglich. | Verfügbarkeit der Konfigurations-UI und API, Testfälle. | - |
| CN2 | Überwachung des Crawling-Prozesses | Die Crawler-Komponente bietet einen geeigneten Überblick über die laufenden Crawling-Prozesse. | Umfrage unter den OPAL-Mitgliedern. | - |
| CN3 | Steuerung des Crawling-Prozesses | Es gibt eine Möglichkeit, die Crawling-Komponente für die OPAL-Administration benutzerfreundlich manuell zu steuern (starten, stoppen). | Umfrage unter OPAL-Mitgliedern. | AK6 |
| CN4 | Doku-mentation der Komponente | Die Entwickler stellen eine Dokumentation für die Konfiguration des Crawlers zur Verfügung. | Dokumentation. | - |
| CN5 | Doku-mentation des Datenflusses | Eine Beschreibung der Speicherung von Daten ist vorhanden. | Dokumentation. | - |
| CN6 | Zeiteffizientes Crawling | Mehrere Instanzen oder Sub-Komponenten ermöglichen das parallele Crawlen einzelner oder mehrerer Ressourcen. | Test von sequentiellen und parallelen Crawling-Jobs. | |
| CN7 | Ordentliches Crawling | Der Crawler sollte robots.txt respektieren und angemessene Pausen zwischen den Zugriffen einfügen. | Integrationstest mit Server-Mock, der das Verhalten des Crawlers verfolgt. | - |

**Vergleich bestehender Crawing-Framworks**

Gemäß der Vorhabensbeschreibung sollte die Implementierung der fokussierten Crawling-Komponente im OPAL-Projekt ein bestehendes Framework wiederverwenden. In der folgenden Tabelle werden bestehende relevante Open-Source-Crawler-Frameworks identifiziert und im Hinblick auf die bestehenden Anforderungen verglichen.

| Software | CF1 | CF2 | CF3 | CF4 | CF5 | CF6 | CF7 | CF8 | CF9 |
|---|---|---|---|---|---|---|---|---|---|
| WebMagic | X (Nur ein Link) | ✓ (nur HTTP) | X | (✓) | (✓) | (✓) | (✓) | (✓) | X |
| StormCrawler | ✓ | ✓ (nur HTTP) | X | X | X | X | X | X | X |
| Apache Nutch | ✓ | ✓ | X | ✓ | (✓) | (✓) | (✓) | X | X |
| REX | ✓ | ✓ (nur HTTP) | X | ✓ | ✓ | ✓ | ✓ | X | X |
| HTTrack | ✓ | ✓ (nur HTTP) | (✓) | X | X | X | X | X | X |
| ldspider | ✓ | ✓ (nur HTTP) | X | ✓ | ✓ | ✓ | ✓ | ✓ | X |
| slug | ✓ | ✓ (nur HTTP) | (✓) | ✓ | ✓ | ✓ | ✓ | ✓ | X |
| Apache Any23 | X (Nur ein Link) | ✓ (nur HTTP) | X | ✓ | ✓ | ✓ | X | X | X |
| Squirrel | (✓) | ✓ | ✓ | (✓) | ✓ | ✓ | ✓ | ✓ | ✓ |
| TDSP | X | X | X | ✓ | ✓ | ✓ | ✓ | ✓ | X |

Entsprechend dieser Analyse wurde das Squirrel-Framework für eine nachfolgende Crawler-Implementierung ausgewählt, da die meisten Anforderungen direkt unterstützt werden und für die übrigen eine Erweiterung erfolgen kann.

**Weiterführende Inhalte**

- D2.1 Spezifikation der Crawler-Komponente (Matthias Wauer, Geraldo de Souza, Adrian Wilke, Afshin Amini): https://github.com/projekt-opal/doc/blob/master/deliverables/OPAL_D2.1_Spezifikation_der_Crawler-Komponente.pdf

### 2.2.2 Arbeitspaket 2.2: Crawling-Komponente

Notwendige Features der in Arbeitspaket 2.1 gewählte Crawling-Komponente Squirrel wurden im Rahmen von OPAL weiterentwickelt. Squirrel wurde fortlaufend eingesetzt, um Daten der Portale mCLOUD, GovData und dem European Data Portal zu beziehen. Außerdem fand nach Einholung der Erlaubnis ein Download offen zugänglicher Daten des MDM Portals statt. Die Squirrel Dokumentation wurde um entsprechnde Anleitungen erweitert.

Die angewendeten metadatenbasierten Crawlingstrategien sowie die Benchmarks sind in den folgenden Arbeitspaketen beschrieben.

**Weiterführende Inhalte**

- Crawler-Software Squirrel: https://github.com/dice-group/Squirrel/releases
- Squirrel Konfiguration: https://github.com/projekt-opal/squirrel-portals-config
- Squirrel Dokumentation; https://dice-group.github.io/squirrel.github.io/
- Software zum MDM Download: https://github.com/projekt-opal/misc/tree/master/mdm-download
- Gecrawlte Rohdaten: https://hobbitdata.informatik.uni-leipzig.de/OPAL/SourceGraphs/
- Gecrawlte Rohdaten ab April 2020: https://hobbitdata.informatik.uni-leipzig.de/OPAL/processed_datasets/

### 2.2.3 Arbeitspaket 2.3: Metadatenbasierte Crawlingstrategien

Squirrel setzt sich aus zwei Hauptkomponenten zusammen: Dem Frontier und Workern. Um eine vollständig erweiterbare Architektur zu erreichen, setzen beide Komponenten auf das Dependency Injektionsmuster, d.h. sie bestehen aus mehreren Modulen, die die einzelnen Funktionalitäten der Komponenten implementieren. Diese Module können in die Komponenten injiziert werden, was das Hinzufügen von weiteren Funktionalitäten ermöglicht. Zur Unterstützung der Injektion von Abhängigkeiten wurde Squirrel auf Basis des Spring-Frameworks implementiert. Die folgende Abbildung veranschaulicht die Architektur:

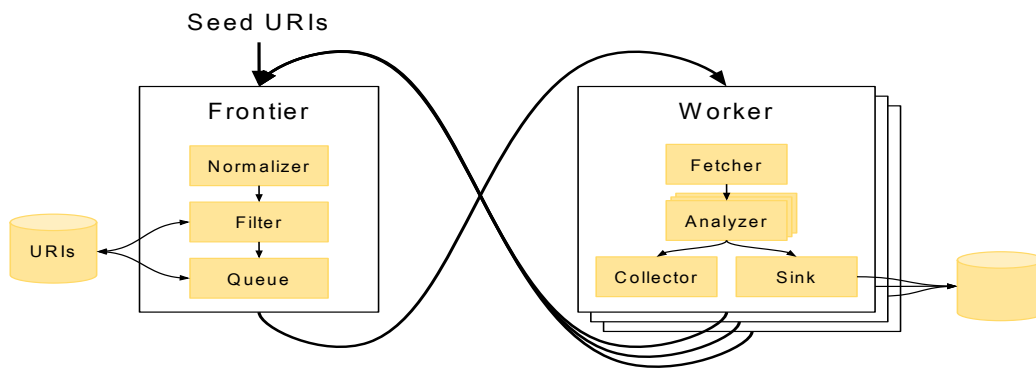

Bei der Ausführung verfügt der Crawler über genau ein Frontier und eine Anzahl von Workern, die parallel arbeiten können. Das Frontier wird mit einer Liste von Seed-URIs initialisiert. Es normalisiert und filtert die URIs. Die beinhaltet die Prüfung beinhaltet, ob die URIs schon einmal durchlaufen wurden. Anschließend werden die URIs in die interne Warteschlange hinzugefügt. Sobald der Frontier eine Anforderung von einem Worker erhält, gibt er einen Satz von URIs an den Worker. Für jede gegebene URI holt der Worker den Inhalt des URIs ab, analysiert die empfangen Daten, sammelt neue URIs und leitet die Daten an seine Senke weiter. Ist der Worker mit dem gegebenen Satz von URIs fertig, sendet er ihn zusammen mit den neu identifizierten URIs zurück an den Frontier Der Crawler implementiert eine periodische Neubewertung von URIs, von denen bekannt ist, dass sie in vergangenen Iterationen gecrawlt wurden.

Die Squirrel Worker dokumentieren den Crawling-Prozess indem sie Metadaten in einen Metadatengraphen schreiben. Diese Metadaten stützen sich hauptsächlich auf der PROV-Ontologie und wurden bei Bedarf erweitert. Die folgende Abbildung gibt einen Überblick über die generierten Metadaten:

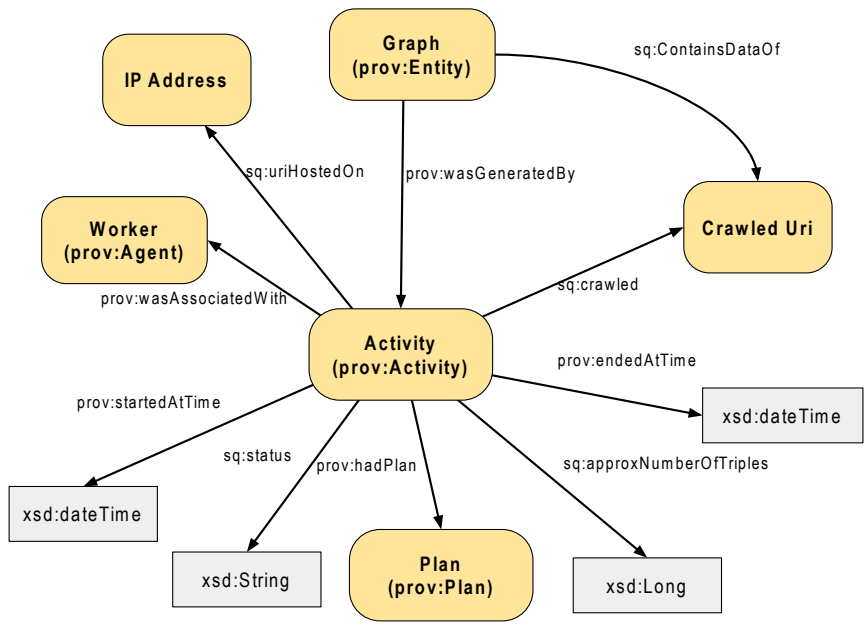

Das Crawlen eines einzelnen URIs wird als eine Aktivität modelliert. Zu einer solchen Aktivität gehören Daten wie die Start- und Endzeit, die ungefähre Anzahl der empfangenen Triples und eine Statuszeile, die anzeigt, ob das Crawling erfolgreich war. Der Ergebnisgraph (oder die Ergebnisdatei im Falle einer dateibasierten Senke) ist eine Entität, die durch die Aktivität erzeugt wird. Sowohl der Ergebnisgraph als auch die Aktivität sind mit dem URI verbunden, der gecrawlt wurde.

Zu den metadatenbasierten wurde ein Artikel und das OPAL Projekt auf der International Semantic Web Conference 2020 vorgestellt und anschließend veröffentlicht.

**Weiterführende Inhalte**

- Squirrel – Crawling RDF Knowledge Graphs on the Web (Artikel von Michael Röder, Geraldo de Souza Jr, Axel-Cyrille Ngonga Ngomo): https://link.springer.com/chapter/10.1007/978-3-030-62466-8_3
- Squirrel – Crawling RDF Knowledge Graphs on the Web (PDF): https://papers.dice-research.org/2020/ISWC_Squirrel/public.pdf

### 2.2.4 Arbeitspaket 2.4: Benchmarking der Crawling-Komponente

Mit diesem Arbeitspaket und dem damit entstandenen ORCA-Benchmark wurde der erste Benchmark entwickelt, um die Leistung von Web und Data Web Crawlern fair zu bewerten. Der Benchmark erzeugt ein synthetisches Daten-Web, das vom Original-Web entkoppelt ist und einen fairen und wiederholbaren Vergleich von Data Web Crawlern ermöglicht. Die Auswertungen zeigen, dass der Benchmark verwendet werden kann, um die verschiedenen Vor- und Nachteile bestehender Crawler aufzuzeigen.

Die Hauptidee hinter ORCA ist die Sicherstellung der vergleichbaren Bewertung von Crawlern zu gewährleisten, indem ein lokales, synthetisches Data Web erstellt wird. Der Benchmark-Crawler wird dabei mit einem Satz von Seed-Knoten der synthetischen Wolke initialisiert und gestartet, um die komplette Wolke zu crawlen. Da die Wolke generiert wird, ist der Benchmark-Komponente bekannt, welche Triples gecrawlt werden sollen und kann die Vollständigkeit und die Geschwindigkeit des Crawlers messen. Da die Generierung der Wolke deterministisch ist, kann eine zuvor verwendete Wolke für das Benchmarking eines anderen Crawlers neu erzeugt werden, so dass dass die Auswertungsergebnisse vergleichbar sind, wenn die Experimente auf der gleichen Hardware ausgeführt werden.

Die folgende Abbildung zeigt eine Übersicht der Benchmark-Komponenten und des Datenflusses. (Orange: Benchmark-Komponenten; Grau: Synthetisches Datennetz; Dunkelblau: Benchmark-Crawler; Durchgehende Linien: Fluss der Daten; Gepunktete Linien: Verknüpfungen zwischen RDF-Datensätzen; die Zahlen geben die Ausführungsreihenfolge an.)

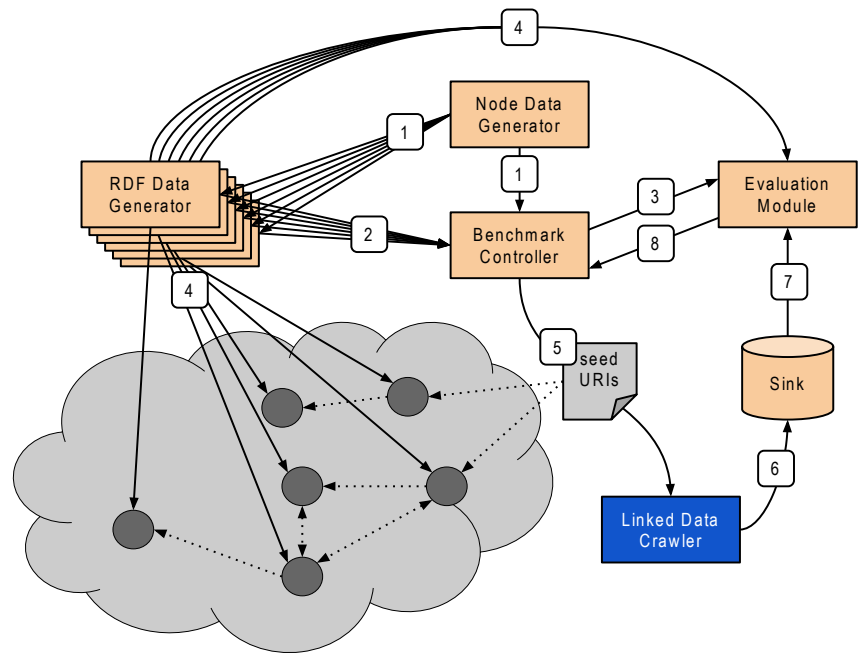

Die Ergebnisse des Benchmarks sind in nachfolgender Abbildung dargestellt. Es wurden der OPAL Crawler Squirrel und der LDSpider Crawler miteinander verglichen. Dabei wurden die Crawler mit unterschiedlichen Konfigurationen gestartet, so etwa Squirrel mit 1, 3, 9 bzw. 18 Worker-Komponenten. Insbesondere für das Data Web lieferte der Squirrel Crawler die besten Ergebnisse. Zudem kann Squirrel im Gegensatz zu LDSpider die RDF/JSON Serialisierung und Kompressionsformate behandeln.

| Crawler | Data Web | | Efficiency | | | |
|---|---|---|---|---|---|---|
| | Micro Recall | Runtime (in s) | Micro Recall | Runtime (in s) | CPU (in s) | RAM (in GB) |
| LDSpider (T8) | 0.00 | 67 | – | – | – | – |
| LDSpider (T16) | 0.00 | 73 | – | – | – | – |
| LDSpider (T32) | 0.00 | 74 | – | – | – | – |
| LDSpider (T1,FS) | 0.31 | 1 798 | 1.00 | 1 847 | 627.2 | 1.8 |
| LDSpider (T8,FS) | 0.30 | 1 792 | 1.00 | 1 717 | 658.9 | 5.2 |
| LDSpider (T16,FS) | 0.31 | 1 858 | 1.00 | 1 753 | 1 677.1 | 1.6 |
| LDSpider (T32,FS) | 0.31 | 1 847 | 1.00 | 1 754 | 1 959.1 | 4.0 |
| LDSpider (T32,FS,LBS) | 0.03 | 66 | 0.01 | 56 | – | – |
| Squirrel (W1) | 0.98 | 6 663 | 1.00 | 12 051 | 1 096.7 | 3.5 |
| Squirrel (W3) | 0.98 | 2 686 | 1.00 | 4 096 | 992.0 | 7.7 |
| Squirrel (W9) | 0.98 | 1 412 | 1.00 | 1 500 | 652.0 | 16.6 |
| Squirrel (W18) | 0.97 | 1 551 | 1.00 | 893 | 424.0 | 24.0 |

Die Benchmarks und das OPAL Projekt wurden als Artikel auf der International Conference on Semantic Computing (ICSC 2021) vorgestellt. Eine Veröffentlichung findet im Laufe des Jahres statt. Die exakte technische Vorgehensweise ist im Artikel / Deliverable 2.6 beschrieben. Die entwickelte Software ist als Open Source veröffentlicht.

**Weiterführende Inhalte**

- ORCA – a Benchmark for Data Web Crawlers (Artikel von Michael Röder, Geraldo de Souza Jr., Denis Kuchelev, Abdelmoneim Amer Desouki, Axel-Cyrille Ngonga Ngomo): https://papers.dice-research.org/2021/ICSC2021_ORCA/ORCA_public.pdf
- D2.3 Benchmark-Spezifikation und Ergebnisse des ersten Crawlers (Geraldo de Souza, Adrian Wilke): https://github.com/projekt-opal/doc/blob/master/deliverables/OPAL_D2.3_Benchmark-specification_and_first_results_Crawler.pdf
- ORCA Software: https://w3id.org/dice-research/orca

## 2.3 Arbeitspaket 3: Datenanalyse

**Ziel des Arbeitspaketes**

Das dritte Arbeitspaket entwickelt Komponenten zur Untersuchung und Gewinnung von Metadaten der in Arbeitspaket 2 gefundenen Information.

**Deliverables des Arbeitspaketes**

Die folgenden Abgaben wurden in diesem Arbeitspaket fertiggestellt:

- D3.1 Spezifikation von Qualitätskriterien
- D3.2 Qualitätsanalyse-Komponente
- D3.3 Erste Metadatenextraktionskomponente
- D3.4 Topic-Extraktionskomponente
- D3.5 Finale Datenanalysekomponenten

### 2.3.1 Arbeitspaket 3.1: Qualitätsanalyse

**Spezifikation von Qualitätskriterien**

In der wissenschaftlichen Literatur wurden bereits unterschiedliche Aspekte zu offenen Daten und Metadaten untersucht. Die hier durchgeführte Literaturrecherche wurde hinsichtlich eines Systems zur automatischen Generierung von Qualitätsmerkmalen von Metadaten zu offenen Datensätzen im Bereich von Mobilitätsdaten durchgeführt. Das Ergebnis der Literaturrecherche ist der folgende Katalog, bestehend aus 13 Qualitätsdimensionen und 48 zugehörigen Qualitätskriterien und -metriken zur konkreten Implementierung.

**Ausdruckskraft**

| Nr. | Kriterium | Beschreibung | Metrik |
|---|---|---|---|
| 1 | Umfang | Ein Metadateneintrag für einen bestimmten Datensatz sollte aus einer umfangreichen, nicht leeren Menge von Metadatenfeldern bestehen. | Zählen der Anzahl der angegebenen Metadatenwerte und teilen dieser durch die Gesamtzahl der verwendeten Metadatenfelder. |
| 2 | Gewichteter Umfang | Für verschiedene Interessensgebiete können bestimmte Metadatenfelder eine zusammenhängende Bedeutung erhalten. Dies kann durch die Anwendung von (benutzerbasierten) Gewichten für einzelne Metadatenfelder quantifiziert werden. | Multiplizieren gegebener Metadatenfelder mit einer Gewichtung, Brechung der Summe aller gewichteten Werte und Teilen durch die größte Summe. |
| 3 | Kategorisierung | Insbesondere Kategorien und Tags können dazu verwendet werden, neue Datensätze zu entdecken. Umgekehrt kann ein aktueller Datensatz aus anderen Quellen entdeckt werden. | Dies kann implementiert werden, indem durch einen true/false-Wert geprüft wird, ob ein Metadatensatz Tags/Kategorien enthält oder nicht. Zusätzlich kann die Anzahl der Kategorien und Tags zur Quantifizierung verwendet werden. |
| 4 | Beschreibung | In Metadaten der realen Welt kann der Beschreibungstext sehr kurz sein oder nur aus einer Kopie des Titels bestehen. | Prüfen, ob der Beschreibungstext nicht leer ist, sich nicht mit dem Titel überschneidet und N Wörter umfasst. |

**Zeitlich**

| Nr. | Kriterium | Beschreibung | Metrik |
|---|---|---|---|
| 5 | Aktualität | Datenbenutzer sind oft an aktuellen Daten interessiert. | Berechnung der Differenz zwischen der aktuellen Zeit und dem Zeitpunkt der Erstellung oder der letzten Aktualisierung. Es wird nur ein Zeitstempel in den Metadaten benötigt. |
| 6 | Aktualisierungs-rate | Informationen über die Häufigkeit der Aktualisierungen. | Summe der Anzahl der Aktualisierungsereignisse. |

**Verständlichkeit**

| Nr. | Kriterium | Beschreibung | Metrik |
|---|---|---|---|
| 7 | Lesbarkeit | Die Lesbarkeit von Texten kann anhand der Satzlänge und der Wortarten geschätzt werden. | Verwendung eines Lesbarkeitstests, z. B. Flesch-Reading-Ease |
| 8 | Sprachfehler | Korrekte Sprache in Metadatenfeldern kann ein Indikator für die Datenqualität sein, insbesondere wenn Teile der Metadaten aus den Originaldaten extrahiert wurden. | Zählen Sie die Anzahl der falsch geschriebenen Wörter. Zusätzlich kann eine Heuristik für Grammatikfehler angewendet werden. |
| 9 | Beispielanwendungen | Beispielanwendungen, die auf Daten basieren, die durch einen Metadatensatz beschrieben werden, können Softwareprojekte oder Artikel sein. | Prüfung, ob es Links zu Beispielen gibt. |

**Rechte**

| Nr. | Kriterium | Beschreibung | Metrik |
|-----|-----------|--------------|--------|
| 10 | Maschinenlesbare Lizenz | Eine maschinenlesbare Lizenz kann automatisch verarbeitet werden, um Informationen für Benutzer bereitzustellen und zu aggregieren. Dazu können IDs wie offizielle Abkürzungen oder URLs von offiziellen Websites verwendet werden. | Prüfung, ob eine ID, URL oder strukturierte Informationen über die Lizenz vorhanden sind. |
| 11 | Von Menschen lesbare Lizenz | In Benutzeroberflächen können aufgelistete Lizenzen dem Benutzer Informationen über Rechte zur Verfügung stellen oder zumindest die Möglichkeit bieten, die zugehörigen Lizenzberechtigungen und -beschränkungen nachzuschlagen. | Prüfung, ob die Metadaten einen Titel der Lizenz enthalten. |
| 12 | Bekannte Lizenz | Eigenschaften bekannter Lizenzen können den Benutzern auf unterschiedliche Weise dargestellt werden und stellen einen Mehrwert dar. | Prüfung, ob die Lizenz in einer Lizenzdatenbank aufgeführt ist. |
| 13 | Offene Lizenz | Offene Lizenzen erweitern typischerweise den Bereich der möglichen Anwendungen für die Arbeit mit den zugehörigen Daten. | Prüfung, ob die Lizenz als offen beschrieben ist. |
| 14 | Erlaubnis zur kommerziellen Nutzung | Die aus den Daten extrahierten Informationen sind wertvoll und können in kommerzielle Datenbanken und Anwendungen integriert oder einfach verkauft werden, um die Arbeit mit den Daten zu honorieren. | Prüfung, ob die Lizenz als frei für kommerzielle Nutzung beschrieben ist. |
| 15 | Erlaubnisse | Einige Datensätze sind eingeschränkt, um den ursprünglichen Autor zu nennen. In anderen Fällen können die Daten für akademische Zwecke verfügbar sein. Das Bewusstsein über diese Einschränkungen ist wichtig für die Wiederveröffentlichung von Daten. | Prüfung, ob die Lizenz Eigenschaften wie Attribution, Share-Alike oder Public-Domain erfüllt. |

**Vertrauen**

| Nr. | Kriterium | Beschreibung | Metrik |
| --- | --- | --- | --- |
| 16 | Provider-Identität | Daten können von Instanzen wie Behörden, Universitäten oder beliebigen Menschen oder Maschinen bereitgestellt werden. Eine bekannte Identität kann für die Recherche von zusätzlichen Informationen verwendet werden. | Prüfung, ob der offizielle Name, Titel oder die Website des Datenanbieters angegeben ist. |
| 17 | Vertrauens-würdiger Provider | Die Glaubwürdigkeit und Zuverlässigkeit des Metadatenanbieters kann in einer Datenbank gespeichert werden und Metadatensätze desselben Anbieters können bewertet werden. | Prüfung, ob der Metadatenanbieter in einer Anbieterdatenbank mit Bewertungsinformationen aufgeführt ist. |
| 18 | Authentizität der Metadaten | Werden mehrere Herausgeber oder Quellen für einen Metadatensatz gefunden oder gibt es Überschneidungen zwischen Datenlieferanten für mehrere Metadatensätze, können diese Informationen zur indirekten Bewertung eines aktuellen Metadatensatzes verwendet werden. | Prüfung des internen Metadatenspeichers auf verwandte Entitäten im Netzwerk und Berechnung eines Scores auf der Basis der verwandten Daten. |
| 19 | Verwendung von digitalen Signaturen | Digitale Signaturen können zur Validierung eines Datenanbieters verwendet werden und somit ein Vertrauensmaß darstellen. | Prüfung der verknüpften Quellen auf Signaturen und Bereitstellung eines booleschen Wertes für die Existenz. |

**Community**

| Nr. | Kriterium | Beschreibung | Metrik |
|-----|-----------|--------------|--------|
| 20 | Kommunikation | Nachvollziehbare Diskussionen über offene Datensätze können das Bewusstsein für den inhaltlichen Zusammenhang schärfen. Außerdem können Online-Kommunikationswerkzeuge aktiv genutzt werden, um zusätzliche Informationen über Daten zu sammeln. | Prüfung, ob Quellen für Online-Diskussionen wie Message Boards oder Mailinglisten angegeben sind. |
| 21 | Vertrauen übertragen | Ein Vertrauenswert, der auf Benutzerabstimmungen basiert, kann Einblicke in die Meinungen anderer bieten. | Prüfung, ob die Metadatenquellen eine benutzerbasierte Vertrauenseinstufung bieten und die entsprechenden Informationen bereitstellen. |
| 22 | Korrektheit | Ein Korrektheitswert, der auf Benutzerstimmen basiert, kann Einblicke in die Meinung anderer geben. | Prüfung, ob die Metadatenquellen ein benutzerbasiertes Korrektheitsranking bereitstellen und die entsprechenden Informationen bereitstellen. |
| 23 | Bestätigung | Metadaten können von verschiedenen Quellen bereitgestellt werden. Eine Überschneidung der gleichen Metadatenwerte kann eine Bestätigung sein, wenn sich die Quellen nicht aufeinander verlassen. | Prüfung, ob mehrere Metadatenquellen die gleichen Metadateninformationen bereitstellen und einen Score aus Überschneidungen berechnen. Wenn sich Metadatenfelder nicht überschneiden, können die unabhängigen Felder aggregiert werden. |

**Vielseitigkeit**

| Nr. | Kriterium | Beschreibung | Metrik |
|-----|-----------|--------------|--------|
| 24 | Mehrere Serialisierungen | Die Metadaten können in verschiedenen Formaten bereitgestellt werden, z. B. XML, JSON oder RDF. Speziell für Mobilitätsdaten könnten z. B. Web Map Service (WMS) und Web Feature Service (WFS) bereitgestellt werden. | Prüfung, ob Links zu verschiedenen Dateiformaten bereitgestellt werden. |
| 25 | Mehrere Sprachen | Metadaten können in verschiedenen Sprachen bereitgestellt werden, insbesondere unstrukturierte Teile wie Beschreibungstexte. | Prüfung, ob Links mit unterschiedlichen Sprachen in URL oder Titel der Links bereitgestellt werden. |
| 26 | Mehrere Zugriffsmethoden | Die gleichen Metadaten können über dateibasierte Repositories, APIs, Abfragesprachen wie SPARQL oder Protokolle wie HTTP oder FTP bereitgestellt werden. | Prüfung, ob Links mit unterschiedlichen Zugriffsmethoden bereitgestellt werden. |

**Repräsentation**

| Nr. | Kriterium | Beschreibung | Metrik |
|-----|-----------|--------------|--------|
| 27 | Offenes Format | Metadaten sollten unter Verwendung eines offenen, nicht-proprietären Standards bereitgestellt werden. | Prüfung, ob das bereitgestellte Metadatenformat Standards des W3C oder ähnlicher Institutionen verwendet. |
| 28 | Datenformat | Das bereitgestellte Format der Daten sollte klar definiert, als offizieller Standard veröffentlicht und bei Institutionen wie IANA registriert sein, um Interoperabilität und Zugriff zu unterstützen. | Prüfung, ob das Metadatenformat mit bekannten Standards konform ist. |
| 29 | Maschinell verarbeitbar | Um eine automatische Verarbeitung zu ermöglichen, sollten die Metadaten entsprechend strukturiert sein. | Prüfung, ob das Metadatenformat eine Struktur unterstützt, die von Maschinen automatisch verarbeitet werden kann. |
| 30 | Vokabular | Metadaten sollten mit einem gut strukturierten, klar definierten Vokabular/Ontologie/Schema beschrieben werden, um vergleichbar, wiederverwendbar und automatisch verarbeitbar zu sein. | Prüfung, ob das verwendete Schema zur Metadatendarstellung angegeben ist. |
| 31 | Datums-Format | Daten sollten in einer standardisierten Form angegeben werden, um Parsing-Fehler und Probleme mit Zeitunterschieden zu vermeiden. | Prüfung, ob die Datumsformate angegeben sind. |
| 32 | Eindeutiger Bezeichner | Zur Identifizierung eines Metadatensatzes sollte ein universeller, eindeutiger und dauerhafter Bezeichner verwendet werden. | Prüfung, ob Informationen über eine UID angegeben sind. |
| 33 | Lokalität | Besonders für Mobilitätsdaten sind Informationen über den geografischen Kontext wichtig. Diese hierarchischen Informationen können oft verfeinert werden, z. B. wenn der Wert der Ortsangabe ein Bundesland ist, das hierarchisch verfeinert werden kann, um auch enthaltene Regionen wie Städte zuzuordnen. | Prüfung, ob es ein Metadatenfeld für die geografische Region gibt oder ob diese Information aus unstrukturierten Texten innerhalb der Metadaten extrahiert werden kann. |

## Verknüpfung

| Nr. | Kriterium | Beschreibung | Metrik |
|---|---|---|---|
| 34 | Gelabelte Daten | Metadatenfelder und Werte in den Feldern werden beschriftet, um zumindest halbstrukturierte Daten bereitzustellen. Beschriftete Entitäten sollten für Menschen und Maschinen lesbar sein. | Prüfung, ob die Daten zumindest halbstrukturiert (z. B. durch Verwendung von XML) oder strukturiert (z. B. durch Verwendung von RDF) sind |
| 35 | Verknüpfte Datendarstellung | Metadaten werden als verknüpfte Daten bereitgestellt und als RDF oder Ansätze auf höheren Ebenen dargestellt. | Verwendung von RDF prüfen |
| 36 | Metadaten-Verknüpfung | Die verwendete Metadaten-Darstellung wird mit Verknüpfungen zu anderen Metadaten-Standards erweitert, um die automatische Verarbeitung und Schlussfolgerung durch Maschinen zu unterstützen. | Prüfung, ob Ontologien/Namensräume/Schemata anderer Standards bereitgestellt werden. |

## Erreichbarkeit

| Nr. | Kriterium | Beschreibung | Metrik |
|---|---|---|---|
| 37 | Kontakt-URL | Eine Kontakt-Website des Datenanbieters ermöglicht die Kontaktaufnahme und das Auffinden verwandter Datensätze desselben Anbieters. | Prüfung, ob eine URL des Anbieters angegeben ist. |
| 38 | Kontakt-E-Mail | Metadatensätze können eine Kontakt-E-Mail-Adresse enthalten, um verantwortliche Personen der Daten zu erreichen und die Option für Datenverbesserungen zu eröffnen. | Prüfung, ob eine Kontakt-E-Mail angegeben ist und das Format gültig ist (z. B. verwendete Zeichen, Verwendung eines At-Zeichens, angegebene Domain). |
| 39 | Klassische Kontakt-informationen | Klassische, nicht-digitale Kontaktinformationen wie Adresse, Telefonnummer, zuständige Abteilung oder Person bieten zusätzliche Möglichkeiten, den Datenanbieter zu kontaktieren. | Prüfung, ob die Metadaten entsprechende Daten enthalten, idealerweise strukturiert durch Felder für Adresse, Telefonnummer und ansprechbaren Akteur. |

## Zugriff

| Nr. | Kriterium | Beschreibung | Metrik |
|---|---|---|---|
| 40 | Offene Metadaten | Auf die Metadaten kann ohne Einschränkungen und ohne Registrierung zugegriffen werden. | Prüfung, ob es eine Möglichkeit gibt, auf die Metadaten offen zuzugreifen. |
| 41 | Wieder-auffindbarkeit | Die Metadaten sollten von einem Agenten abgerufen werden und die Antwort sollte einen Code für Erfolg zurückgeben (z. B. HTTP 200 oder FTP 2xx) | Prüfung, ob auf die Metadaten zugegriffen werden kann und ein Erfolgscode zurückgegeben wird. |

## Versionierung

| Nr. | Kriterium | Beschreibung | Metrik |
|---|---|---|---|
| 42 | Versions-nummerierung | Für verschiedene Zustände von Daten kann in den Metadaten eine Versionsnummer angegeben werden, um eine Kennung für die aktuelle Version zu liefern. | Prüfung, ob die Metadaten ein eigenes Feld für die Version enthalten oder versuchen, diese aus den Beschreibungstexten in den Metadaten zu extrahieren. |
| 43 | Zeitspanne | Die Sammlung, Erzeugung oder Aggregation von Daten erfolgt zu einem bestimmten Zeitpunkt. Dies kann eine wertvolle Information für Datenbenutzer sein. | Prüfung, ob ein zugehöriges Feld und ein Wert Teil des Metadatensatzes ist. Zusätzlich können unstrukturierte Texte auf diese Informationen geprüft werden. |

## Daten

| Nr. | Kriterium | Beschreibung | Metrik |
|---|---|---|---|
| 44 | Offenes Datenformat | Die Daten sollten in einem offenen, nicht-proprietären Standard bereitgestellt werden. Dies kann durch Informationen in den Metadaten überprüft werden. | Prüfung, ob das bereitgestellte Datenformat Standards vom W3C oder ähnlichen Institutionen verwendet. |
| 45 | Datenformat | Das bereitgestellte Datenformat sollte klar definiert, als offizieller Standard veröffentlicht und bei Institutionen wie IANA registriert sein, um Interoperabilität und Zugriff zu unterstützen. Dies kann anhand der Informationen in den Metadaten überprüft werden. | Prüfung, ob das Datenformat mit bekannten Standards konform ist. |
| 46 | Maschinenver-arbeitbare Daten | Um eine automatische Verarbeitung zu ermöglichen, sollten die Daten entsprechend strukturiert sein. Dies kann durch Informationen in den Metadaten überprüft werden und ist eine Spezialisierung der allgemeineren Sicht auf Datenformate. | Prüfung, ob das Datenformat eine Struktur unterstützt, die von Maschinen automatisch verarbeitet werden kann. |
| 47 | Eindeutiger Datenbezeichner | Um einen Datensatz zu identifizieren, sollte ein universeller, eindeutiger und dauerhafter Bezeichner verwendet werden. Diese Information kann in den Metadaten angegeben werden. | Prüfung, ob Informationen über eine Daten-UID gegeben sind. |
| 48 | Mehrere Daten-serialisierungen | Die Daten können in verschiedenen Formaten bereitgestellt werden, z. B. XML, JSON oder RDF. Diese können innerhalb der Metadaten verlinkt werden. | Prüfung, ob Links zu verschiedenen Dateiformaten angegeben sind. |

**Qualitätsanalyse-Komponente**

Mit der Qualitätsanalyse-Komponente erfolgte eine Implementierung von Qualitätsmetriken. Die Komponente nimmt als Eingabe einen RDF-Graphen entgegen, der Tripel eines Metadateneintrages im DCAT Format enthalten muss. Vor Ausführung der Komponente besteht die Möglichkeit, folgende Einstellungen vorzunehmen:

- setIncludeLongRunning: Entscheidung, ob Metriken berechnet werden sollen, die voraussichtlich eine Lange Laufzeit benötigen. Dies ist z.B. die RetrievabilityMetric, bei deren Ausführung getestet wird, ob URIs im Web zugreifbar sind. In der Ausgangskonfiguration ist diese Option nicht gesetzt.
- setLogIfNotComputed: Entscheidung, ob in einer Datei festgehalten werden soll, falls eine Metrik für einen Datensatz z.B. wegen eines Ausnahmefehlers nicht berechnet werden kann. Diese Option ist in der Ausgangskonfiguration gesetzt.
- setRemoveMeasurements: Entscheidung, ob etwaige bereits bestehende Ergebnisse entfernt werden sollen. Dieser Fall trifft z.B. ein, wenn die Komponente mehrfach für denselben Datensatz ausgeführt wird. In der Ausgangskonfiguration ist diese Option gesetzt.

Als Ausgabe werden Ganzzahlen im Intervall [0..5] geschrieben, wobei 0 das schlechteste und 5 das beste Ergebnis darstellt. Als Vokabular wird das Data Quality Vocabulary (DQV) verwendet. Die folgende Abbildung zeigt ein Minimalbeispiel der Daten einer Ausführung:

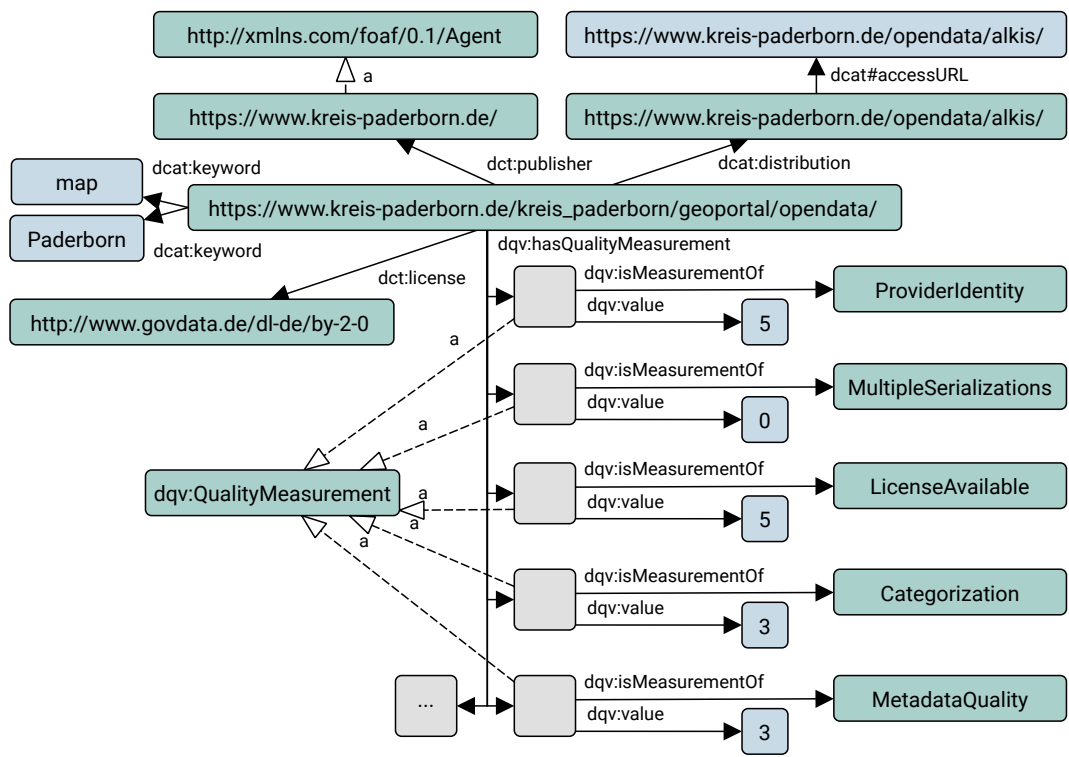

Im Beispiel wurde als Eingabe (oberer Teil) eine Ressource genutzt, für die Angaben zur Lizenz, Stichwörtern, Herausgeber und einer Datenversion bestehen. Als Resultat wurden Werte für 5 Metriken berechnet. Die Komponente ist unter dem Namen *Civet* als Open Source veröffentlicht.

**Weiterführende Inhalte**

- D3.1 Spezifikation von Qualitätskriterien (Adrian Wilke, Ivan Ermilov, Matthias Wauer): https://github.com/projekt-opal/doc/blob/master/deliverables/OPAL_D3.1_Specification_of_quality_criteria.pdf
- D3.2 Qualitätsanalyse-Komponente (Adrian Wilke, Matthias Wauer): https://github.com/projekt-opal/doc/blob/master/deliverables/OPAL_D3.2_Quality-analysis-component.pdf
- Software Qualitätsanalyse: https://github.com/projekt-opal/civet

### 2.3.2 Arbeitspaket 3.2: Semistrukturierte Metadatenextraktion

**Metadatenextraktion aus Portalen**

Semistrukturierte Metadaten sind in unterschiedlichen Datenrepräsentationen aufzufinden, beispielsweise in HTML-Tabellen. Durch maschinelle Erstellung solcher Tabellen tauchen innerhalb der Datenstrukturen Artefakte auf, die zur Identifizierung und Extraktion von Nutzdaten verwendet werden können. In der Anfangsphase des OPAL-Projekts wurden CSS-Klassenbezeichner genutzt, um den Squirrel-Crawler zu konfigurieren und relevante Daten zu extrahieren. Dieses Vorgehen ist während der Projektlaufzeit obsolet geworden, da sich Aktualisierungen der Portale ergaben. Beispielsweise wurde mit dem mCLOUD Release 1.5.0 eine maschinenlesbare Schnittstelle eingeführt, womit strukturierte Metadaten verfügbar gemacht wurden. Die ursprünglich verwendete Konfiguration des Squirrel Crawlers ist weiterhin in der Code-History auffindbar.

**Metadatenextraktion von Lizenzen**

Für den Bericht zur ersten Metadatenextraktionskomponente (Deliverable 3.3) wurden zudem eine Softwarekomponente geringen Umfangs zur Extraktion von Lizenz-URIs erstellt. Für das Abgabedokument wurden Lizenzen von rund 2.800 Metadatensätzen und 61.000 Distributionen (Versionen) aus mCLOUD und dem European Data Portal extrahiert. Die meistverwendete Lizenz dieser Statistik ist die Open Government Licence v2.0.

Im weiteren Projektverlauf wurden Lizenz-Identifikatoren (URIs) zur Referenzierung von Lizenzen genutzt und mit der ReCoDa-Komponente eine Lösung zur Berechnung von Kompatibilitäten umgesetzt (siehe Arbeitspaket 5.4).

**Weiterführende Inhalte**

- D3.3 Erste Metadatenextraktionskomponente (Adrian Wilke, Michael Röder): https://github.com/projekt-opal/doc/blob/master/deliverables/OPAL_D3.3_Metadaten-Extraktion.pdf
- Squirrel-Konfiguration für mCLOUD: https://github.com/projekt-opal/squirrel-portals-config/commits/master/mcloud.yaml
- mCLOUD Release 1.5.0: https://mcloud.de/web/guest/blog/-/blogs/mcloud-release-1-5-0
- Software Daten-Statistiken: https://github.com/projekt-opal/statistics

### 2.3.3 Arbeitspaket 3.3: Unstrukturierte Metadatenextraktion

**Spracherkennung**

In RDF besteht die technische Möglichkeit, Literale mit einer Sprachauszeichnung zu versehen. Für Literale, zu denen keine Sprache spezifiziert ist, wird eine Spracherkennung auf Basis von Apache OpenNLP durchgeführt. Als Eingabe werden hier Titel- und Beschreibungstext von Metadatensätzen verwendet. Im Fall dass Apache OpenNLP für eine Sprache eine Wahrscheinlichkeit oberhalb eines Schwellenwertes zurück gibt, werden die entsprechenden Sprach-Tags von Titel- und Beschreibungsliteralen aktualisiert.

**Geografische Daten**

Zur Anreicherung der Metadatensätze mit geografischen Angaben wurden die Geodatenbanken Nomenclature of Territorial Units for Statistics (NUTS) und Local Administrative Units (LAU) verwendet. Mit diesen steht eine hierarchische Datenbank mit Geopolygonen und entsprechenden Eigennamen von Orten zur Verfügung. Für die Datenextraktion wurde zunächst die Komponente LauNuts verwendet. Auf dieser Basis können 8.495 Orte in Deutschland in Volltexten detektiert werden. Die angereicherten Daten werden im Projektverlauf für eine Geo-Suche verwendet.

**Eigennamenerkennung und Disambiguierung von Entitäten**

Zur Named Entity Recognition wurde FOX (Federated knOwledge eXtraction framework) eingesetzt. Die eingebettete Komponente kam für die Eigennamenerkennung von Orten zum Einsatz. Hierzu wurden folgenden Webservices implementiert:

- metadata/fox: Gibt die extrahierten Ergebnisse aus FOX im Turtle-Format zurück.
- metadata/fox/location/names: Gibt erkannte Ortsnamen im JSON-Format zurück.
- metadata/fox/location/uris: Gibt Identifikatoren (URIs) erkannter Ortsnamen im JSON-Format zurück.

Während das FOX-Framework selbst die Eigennamenerkennung übernimmt, nutzt das Framework AGDISTIS/MAG zur Disambiguierung der Entitäten. Siehe hierzu AP 4.4: Indizierungskomponente.

**Weiterführende Inhalte**

- D3.3 Erste Metadatenextraktionskomponente (Adrian Wilke, Michael Röder): https://github.com /projekt-opal/doc/blob/master/deliverables/OPAL_D3.3_Metadaten-Extraktion.pdf
- Software Metadatenextraktion: https://github.com/projekt-opal/metadata-refinement
- Software Metadatenextraktion (mit Named Entity Recognition): https://github.com/projekt-opal/metadata-refinement/tree/metadata-alpha
- Software LauNuts: https://github.com/projekt-opal/LauNuts
- Software FOX - Federated Knowledge Extraction Framework: https://github.com/dice-group/FOX

### 2.3.4 Arbeitspaket 3.4: Topic- und Schema-Extraktion

**Topic-Extraction**

Zur Extraktion von Entitäten aus englischsprachigen Texten wurde in Zusammenarbeit mit dem LIM-BO Projekt eine Komponente basierend auf dem Maschinellen Lernen mittels der Bibliothek Rasa NLU (Natural Language Understanding) entwickelt. Basierend auf vorhandenen Beschreibungstexten wurde ein Modell trainiert um Orte, Stichwörter, Daten und Personen zu erkennen. Dadurch dass die Ursprungsdaten von unterschiedlichen Herausgebern und Autoren stammen und die Texte daher einen heterogenen Aufbau mit sich bringen, wurde im weiteren Projektverlauf die Extraktion von Orten mittels LauNuts fokussiert (siehe Arbeitspaket 3.3: Unstrukturierte Metadatenextraktion). Die folgende Tabelle zeigt die Ergebnisse der Topic-Komponente:

|                  | Genauigkeit (Precision) | Sensitivität (Recall) | F-Maß  |
|------------------|-------------------------|-----------------------|--------|
| Datum            | 1.0000                  | 0.0870                | 0.1600 |
| Stichwort        | 0.0000                  | 0.0000                | 0.0000 |
| Person           | 0.0000                  | 0.0000                | 0.0000 |
| Ort              | 1.0000                  | 0.3235                | 0.4889 |
| Keine Entität    | 0.8974                  | 1.0000                | 0.9459 |
| Mittel Micro     | 0.8990                  | 0.8990                | 0.8990 |
| Mittel Makro     | 0.5795                  | 0.2821                | 0.3190 |
| Mittel gewichtet | 0.8626                  | 0.8990                | 0.8602 |

**Klassifizierung von Kategorien**

Für Metadatensätze, die Beschreibungstexte beinhalten, jedoch nicht kategorisiert sind, wurde eine Komponente zur automatischen Klassifizierung von Kategorien entwickelt. Als Kategorien wurden Vorgaben verwendet, die im European Data Portal bestehen. Unter Verwendung der WEKA-Bibliothek wurden TF-IFD Wortvektoren generiert sowie J48-Entscheidungsbäume und ein Naiver Bayes-Klassifikator angewendet. Die Software wurde im Rahmen des OPAL Open Data Hackathon erstellt. Die folgende Tabelle zeigt Ergebnisse einer Kreuzvalidierung eines Durchlaufs mit 160 Instanzen:

| Klassifizierer | 1-Gram  | 2-Gram  | 3-Gram  | 4-Gram  |
|----------------|---------|---------|---------|---------|
| J48            | 75,625% | 59,375% | 59,375% | 59,375% |
| NaiveBayes     | 47,5%   | 31,875% | 36,875% | 35%     |

**Weiterführende Inhalte**

- Software Topic-Extraction: https://github.com/projekt-opal/Topic-Extraction
- Software Classification: https://github.com/projekt-opal/classification
- Rasa NLU: https://pypi.org/project/rasa-nlu/
- WEKA: https://www.cs.waikato.ac.nz/ml/weka/

## 2.4 Arbeitspaket 4: Datenkonvertierung

**Ziel des Arbeitspaketes**

Die extrahierten Metadaten werden in für Linked-Data-Anwendungen geeignete Formate konvertiert. Für die effiziente Durchsuchbarkeit werden geeignete Indexstrukturen entwickelt.

**Deliverables des Arbeitspaketes**

Die folgenden Abgaben wurden in diesem Arbeitspaket fertiggestellt:

- D4.1 Vokabularspezifikation
- D4.2 Konvertierungskomponente
- D4.3 Prototyp Indexstrukturen und Entitätserkennung
- D4.4 Indizierungskomponente

### 2.4.1 Arbeitspaket 4.1: Vokabularspezifikation

Nach Sichtung von Literatur, Portalen und verfügbaren Schnittstellen wurde 2018 als Hauptvokabular eine Auswahl von Identifikatoren aus dem Data Catalog Vocabulary (DCAT) in Version 1 festgelegt. Nachdem im Februar 2010 DCAT Version 2 als W3C Recommendation veröffentlicht wurde, fand eine Erweiterung des OPAL Vokabulars auf dies statt. DCAT spezifiziert Klassen und deren Beziehungen und gibt Empfehlungen für den Einsatz weiterer Vokabularien (z.B. Dublin Core, Friend of a Friend), die in OPAL berücksichtigt und verwendet werden. Für Qualitätsmetriken wird das Data Quality Vocabulary (DQV) eingesetzt (siehe Beispiel in AP 3.1: Qualitätsanalyse). In Fällen in denen es notwendig war eigene Identifizierer zu spezifizieren wurden die entsprechenden Konstanten in der OPAL Common Komponente festgelegt. Die folgende Abbildung zeigt eine Übersicht der wichtigsten Klassen und Beziehungen aus DCAT und DQV, so wie sie in OPAL Verwendung finden. Dabei stellt Dataset eine Spezialisierung von Resource dar, so dass alle Prädikate auch dort verwendet werden. Ein Dataset beschreibt einen Metadateneintrag und Distribution dessen Versionen, die beispielsweise als verschiedene Repräsentationsformen (z.B. CSV und Excel) umgesetzt werden.

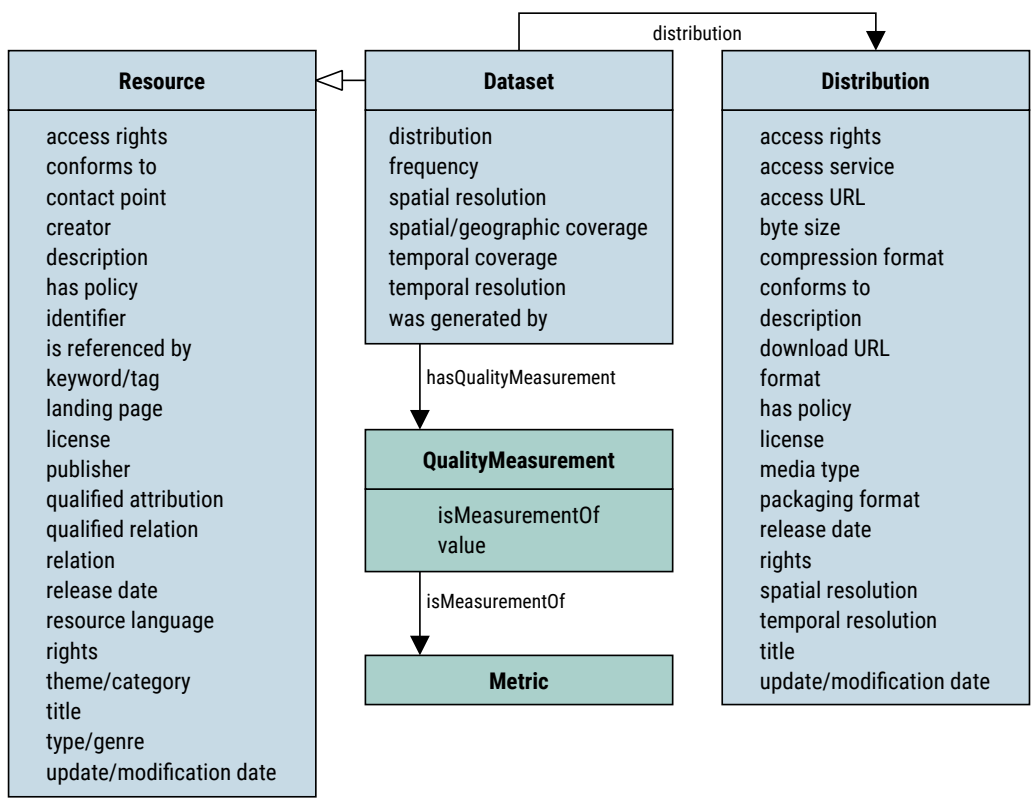

### Weiterführende Inhalte

- D4.1 Vokabularspezifikation (Afshin Amini, Zafar Habeeb Syed): https://github.com/projekt-opal/doc/blob/master/deliverables/OPAL_D4.1_Vocabulary-specification.pdf
- Data Catalog Vocabulary (DCAT): www.w3.org/TR/vocab-dcat-2
- Data Quality Vocabulary (DQV): www.w3.org/TR/vocab-dqv
- OPAL Common Vokabular: https://github.com/projekt-opal/common/tree/master/src/main/java/org/dice_research/opal/common/vocabulary

### 2.4.2 Arbeitspaket 4.2: Konvertierungskomponente

Im Projektverlauf wurde die ursprünglich geplante Architektur mehrfach erweitert und entsprechende Konvertierungsansätze getestet und angepasst. In der ursprünglich festgesetzten Architektur wurde ein Triplestore in der Datenhaltungsschicht zur Weitergabe von Datenversionen verwendet. Da sich dies als Flaschenhals herausstellte, erfolgte eine Erweiterung um Elasticsearch. Später wurde die auf Microservices aufbauende Converter Software auf den finalen OPAL Batch Ansatz umgestellt.

**Konvertierungskomponente Converter**

Die Converter-Komponente baut auf Publish-Subscribe Microservices über Spring Cloud auf. In Warteschlangen zur Verarbeitung werden Datensätze von mehreren, auf jeweilige Anwendungsfälle spezialisierte, Diensten verarbeitet. Änderungen in diesem funktionierenden Verfahren stellte sich als aufwändig heraus, so dass später ein Umstieg auf das im Folgenden aufgeführte Batch-Verfahren erfolgte. Die nachfolgende Abbildung zeigt schematisch den Ablauf der ursprünglichen Verarbeitungskette der Converter-Komponente:

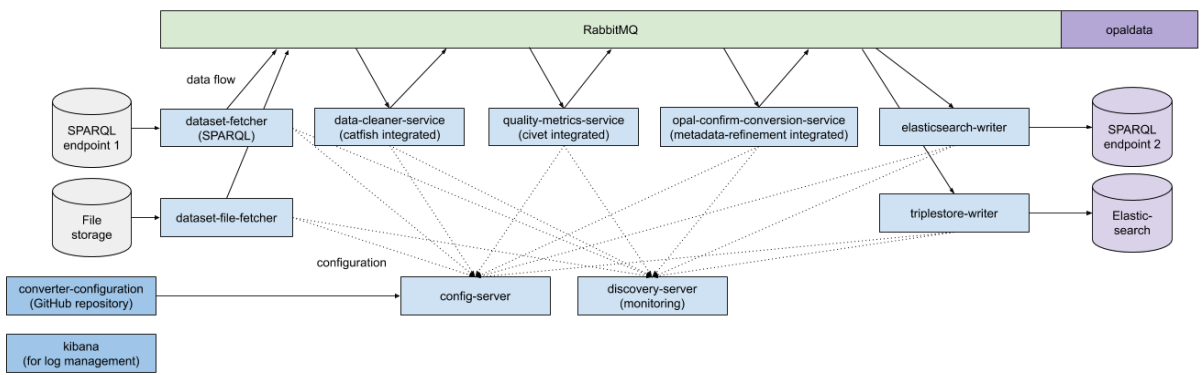

**Konvertierungskomponente Batch**

Die finale Konvertierungskomponente Batch integriert das Lesen von gecrawlten Ursprungsgraphdateien, die Ausführung einzelner Komponenten und das Schreiben nach Elasticsearch sowie ins Dateisystem. Die Konfiguration geschieht über eine Textdatei im Java-Properties Format, in der Einstellungen zu den einzelnen Berechnungsschritten festgelegt werden können. Nach dem Lesen der RDF-Graphen erfolgt eine Unterteilung in Datensatz-Graphen für jeweils ein DCAT Dataset und anschließend eine sequentielle Behandlung einzelner Datensatz-Graphen. OPAL Batch führt die folgenden Einzelkomponenten aus:

- Catfish: In der Datenbereinigung werden, je nach Konfiguration, inhaltsleere Strukturen entfernt, Tripel nicht-spezifizierter Sprachen entfernt, Datenformate bereinigt und Katalog-Identifikatoren gesetzt.
- Metadata-Refinement: Es erfolgt eine Spracherkennung und eine Anreicherung mit Geodaten (siehe AP 3.3: Unstrukturierte Metadatenextraktion).
- Civet: Die Metadatenqualität der Datensätze wird geprüft und bewertet (siehe AP 3.1: Qualitätsanalyse).

Die folgende Abbildung zeigt den Datenfluss der Konvertierungskomponente:

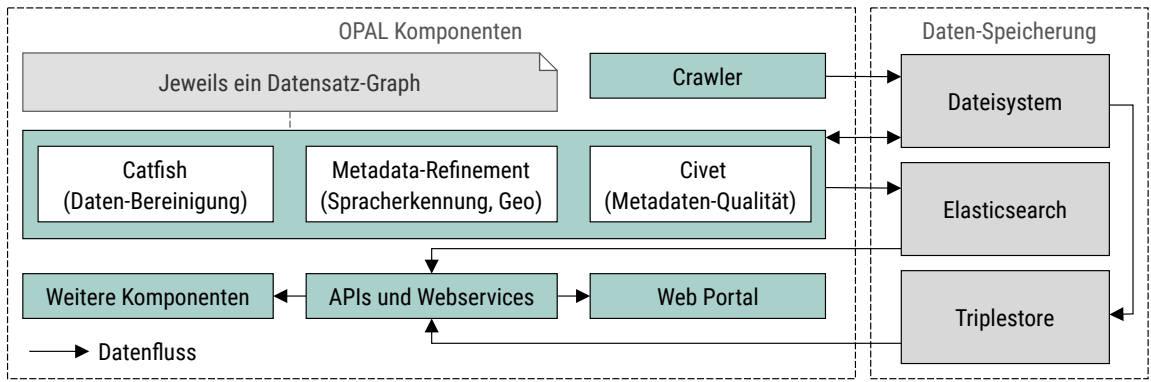

**Weiterführende Inhalte**

- D4.2 Konvertierungskomponente (Afshin Amini, Zafar Habeed Syed, Matthias Wauer): https://github.com/projekt-opal/doc/blob/master/deliverables/OPAL_D4.2_Conversion_component.pdf
- Software Batch: https://github.com/projekt-opal/batch
- Software Catfish: https://github.com/projekt-opal/catfish
- Software Converter: https://github.com/projekt-opal/converter
- Konfiguration Converter: https://github.com/projekt-opal/converter-configuration

### 2.4.3 Arbeitspaket 4.3: Indexstrukturen

Mit der Einführung von Elasticsearch für einen performanten Datenzugriff der Suchkomponente (siehe AP 7.1: Suche) für die Weboberfläche des OPAL-Portals wurde eine Konfiguration der Indexstrukturen erstellt. Primär ist dies ein Mapping der verfügbaren DCAT-Ressourcen und Properties für Elasticsearch. Die entsprechenden typisierten Felder für deutsch- und englischsprachige Texte sowie eingebettete Felder befinden sich in der Mappings Datei in der Datenhaltungskomponente. In Deliverable 4.3 wurde zudem ein erster Benchmark durchgeführt, der die Geschwindigkeitsvorteile von Elasticsearch gegenüber dem Triplestore Virtuoso bestätigt.

**Weiterführende Inhalte**

- D4.3 Prototyp Indexstrukturen und Entitätserkennung (Caglar Demir, Adrian Wilke): https://github.com/projekt-opal/doc/blob/master/deliverables/OPAL_D4.3_Index-structures.pdf
- Elasticsearch Mappings: https://github.com/projekt-opal/opaldata/blob/master/elasticsearch-initialization/mappings.json

### 2.4.4 Arbeitspaket 4.4: Indizierungskomponente

Als Indizierungskomponente wurde die Software AGDISTIS in Kooperation mit dem LIMBO-Projekt erweitert. AGDISTIS wurde um Mehrsprachigkeit (MAG) sowie eine Elasticsearch-Anbindung erweitert und zur Disambiguierung/Linking von Entitäten eingesetzt. Es handelt sich um ein Graph-basiertes Verfahren, in dem der HITS Algorithmus eingesetzt wird. Die Integration in OPAL fand über die Einbindung der geografischen Datenbank LauNuts (siehe AP 3.3: Unstrukturierte Metadatenextraktion), der Eigennamenerkennung (siehe ebenfalls AP 3.3) und der Question-Answering Lösung (siehe AP 7.3: Social Bot) statt.

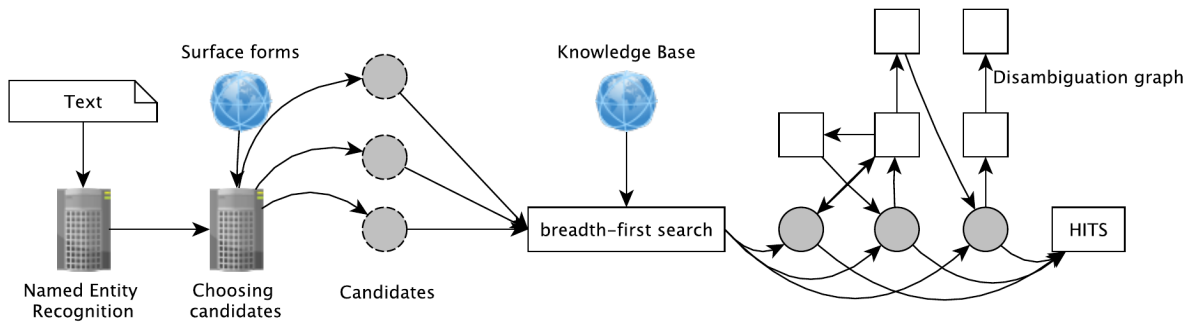

Die neuen wissenschaftlichen Ansätze sowie das OPAL Projekt wurden u.a. auf den Konferenzen European Semantic Web Conference (ESWC 2018) und Knowledge Capture Conference (K-CAP 2017) vorgestellt.

**Weiterführende Inhalte**

- Software Indizierungskomponente AGDISTIS: https://github.com/projekt-opal/AGDISTIS
- Entity Linking in 40 Languages using MAG (Artikel von Diego Moussallem, Ricardo Usbeck, Michael Röder, Axel-Cyrille Ngonga Ngomo): https://arxiv.org/abs/1805.11467
- MAG: A Multilingual, Knowledge-base Agnostic and Deterministic Entity Linking Approach (Artikel von Diego Moussallem, Ricardo Usbeck, Michael Röder, Axel-Cyrille Ngonga Ngomo): https://svn.aksw.org/papers/2017/KCAP_MAG/public.pdf

## 2.5 Arbeitspaket 5: Datenintegration

**Ziel des Arbeitspaketes**

Metadaten verschiedener Datensätze sollen automatisiert verknüpft werden, um miteinander in Relation stehende Daten zu erkennen. Damit soll OPAL es ermöglichen, dass Nutzer für komplexere Anwendungsfälle die dafür geeignete Menge an Datensätzen auffinden.

**Deliverables des Arbeitspaketes**

Die folgenden Abgaben wurden in diesem Arbeitspaket fertiggestellt:

- D5.1 Erste Version der Verknüpfungskomponente
- D5.2 Lernalgorithmen für Linkspezifikationen auf Metadaten
- D5.3 Lizenzintegrationskomponente
- D5.4 Erweiterte Lernalgorithmen für Linkspezifikationen auf Metadaten
- D5.5 Finale Verknüpfungskomponente

### 2.5.1 Arbeitspaket 5.1: Datenverknüpfung

Als Komponente zur Datenverknüpfung wurde LIMES (Link Discovery Framework for Metric Spaces) ausgewählt und im Verlauf des Projekts erweitert. LIMES kann dazu verwendet werden, Überschneidungen in verschiedenen Wissensgraphen zu finden und wurde in OPAL primär eingesetzt, um Verknüpfungen zwischen gleichen Ressourcen in verschiedenen Wissensgraphen zu erstellen. Der Fokus liegt dabei auf geografischen Ähnlichkeiten, die durch Einsatz unterschiedlicher Metriken gefunden werden. Der ursprüngliche LIMES-Ansatz verwendet die Dreiecksungleichung um die Anzahl benötigter Vergleiche zwischen Wissensgraphen zu verringern.

Die Erweiterung WOMBAT, ein Ansatz des Maschinellen Lernens (ML) zur Link Discovery, wurde ebenso eingesetzt. Der Ansatz besteht aus zwei aufeinander aufbauenden Teilen: Ein einzelner (atomic) Vergleich von Bestandteilen zweier Graphen und eine anschließende Kombinierung (generalization). Hierbei bestehen die Möglichkeiten, das Supervised Learning, das ausschließlich positive Lernbeispiele benötigt, einzusetzen oder das Unsupervised Learning, das ohne Lernbeispiele auskommt.

Im Rahmen von OPAL wurde auch DRAGON (Decision Tree Learning for Link Discovery) entwickelt, dass Entscheidungsbäume zum Erlernen von Linkspezifikationen verwendet und beim rekursiven Erstellen von Entscheidungsbäumen entweder die Trefferquote (recall) unter Verwendung des lokalen Gini Index oder die Genauigkeit (precision) unter Verwendung des globalen F-Maß fokussiert.

LIMES wurde im Projektrahmen zu einem Framework erweitert, so auch die LIMES Benutzeroberfläche. Weiterhin kann das RDF Dataset Enrichment Framework DEER verwendet werden. Experimente sind in AP 5.2: Lernalgorithmen für Linkspezifikationen aufgeführt.

**Weiterführende Inhalte**

- Software LIMES - Link Discovery Framework for Metric Spaces: https://github.com/dice-group/LIMES/
- Software DEER - RDF Dataset Enrichment Framework: https://github.com/dice-group/deer
- LIMES - A Framework for Link Discovery on the Semantic Web (Artikel von Axel-Cyrille Ngonga Ngomo, Mohamed Ahmed Sherif, Kleanthi Georgala, Mofeed Hassan, Kevin Dreßler, Klaus Lyko, Daniel Obraczka, Tommaso Soru): https://papers.dice-research.org/2021/KI_LIMES/public.pdf
- Limes Web UI (Artikel von Mohamed Ahmed Sherif, Pestryakova Svetlana, Kevin Dreßler, Axel-Cyrille Ngonga Ngomo): http://ceur-ws.org/Vol-2456/paper53.pdf
- Dragon: Decision Tree Learning for Link Discovery (Artikel von Daniel Obraczka, Axel-Cyrille Ngonga Ngomo)https://doi.org/10.1007/978-3-030-19274-7_51

### 2.5.2 Arbeitspaket 5.2: Lernalgorithmen für Linkspezifikationen

In diesem Teil wurden die Softwarekomponenten aus AP 5.1 in 5 Experimenten zur Deduplizierung angewendet. Aufbauend auf Linkspezifikationen für OPAL/DCAT Datensätzen stand für jedes Experiment jeweils 30 GB Arbeitsspeicher zur Verfügung. Die Experimente sind im Folgenden aufsteigend sortiert nach der Güte ihrer Ergebnisse beschrieben.

- In Experiment 1: WOMBAT als unsupervised complete wurden als Linkspezifikation dcat:Dataset und 23 optionale weitere Eigenschaften festgelegt. Sowohl bei Schwellenwerten von 0,95 und 0,9 wurden keine Duplikate erkannt, da die Ausführung aufgrund der Größe der Eingabedaten bei der trotz 30 GB Arbeitsspeicher mit einem OutOfMemoryError (GC overhead limit exceeded) vorzeitig beendet wurde.
- In Experiment 2: WOMBAT unsupervised simple II wurden ebenfalls dcat:Datasets und 23 optionale weitere Eigenschaften spezifiziert. Hier ergab ein Schwellenwert von 0,9 keine Duplikate und ein Schwellenwert von 0,8 rund 24 Millionen Duplikate. Entsprechend der Ergebniszahlen (keine Ergebnisse und zu viele Duplikate) lieferte dieses Experiment keine verwertbaren Ergebnisse.
- In Experiment 3: WOMBAT unsupervised simple I wurden sowohl dcat:Dataset als auch entsprechende Distributionen dcat:downloadURL und deren 10 optionale weitere Eigenschaften spezifiziert. Ein Schwellenwert von 0,5 ergab rund 18.000 Duplikate, ein Schwellenwert von 0,9 ergab 888 duplizierte Datensätze. Die erkannten 888 Datensätze konnten später bestätigt werden, so dass über Download-URls der jeweiligen Distributionen eine Deduplizierung möglich ist.
- In Experiment 4: LIMES (Dataset) wurde dieselbe Linkspezifikation verwendet; hier wurde LIMES core anstelle von WOMBAT verwendet. Erneut wurden 888 duplizierte Datensätze gefunden.
- In Experiment 5: LIMES (Distribution) wurde die Spezifizierung der Datensätze entfernt und ein Fokus auf Distributionen gesetzt. Es wurden 1.833 Duplikate von Distributionen gefunden.

Abschließend empfehlen wir zum Finden von Duplikaten die Verwendung von Download-URLs, um mehrfach auftretende Datensätze zu detektieren. Dies ist für Anwendungsfälle relevant, in denen Daten aus mehreren Portalen zur Verfügung stehen und gleiche Datensätze erkannt werden müssen.

**Weiterführende Inhalte**

- D5.2 Lernalgorithmen für Linkspezifikationen auf Metadaten (Adrian Wilke, Mohamed Sherif): OPAL Datenintegration
- Software zur Datenextraktion: OPAL Datenintegration

### 2.5.3  Arbeitspaket 5.3: Erweiterte Lernalgorithmen über Daten

Die Umsetzung von erweiterte Lernalgorithmen fand in ORCHID - Reduction-Ratio-Optimal Computation of Geo-Spatial Distances for Link Discovery statt. Hierzu wurden Experimente zum Linking von Geodaten durchgeführt. Eingesetzt wurden primär die Datenbank OPAL LauNuts mit 84.000 Koordinaten-Punkten und LinkedGeoData, ein OpenStreetMap Derivat im RDF-Format. Die Distanzen der Geoobjekte konnten in performanter Weise errechnet werden. Im Folgenden sind die Laufzeiten von synthetischen und realen Daten aufgeführt:

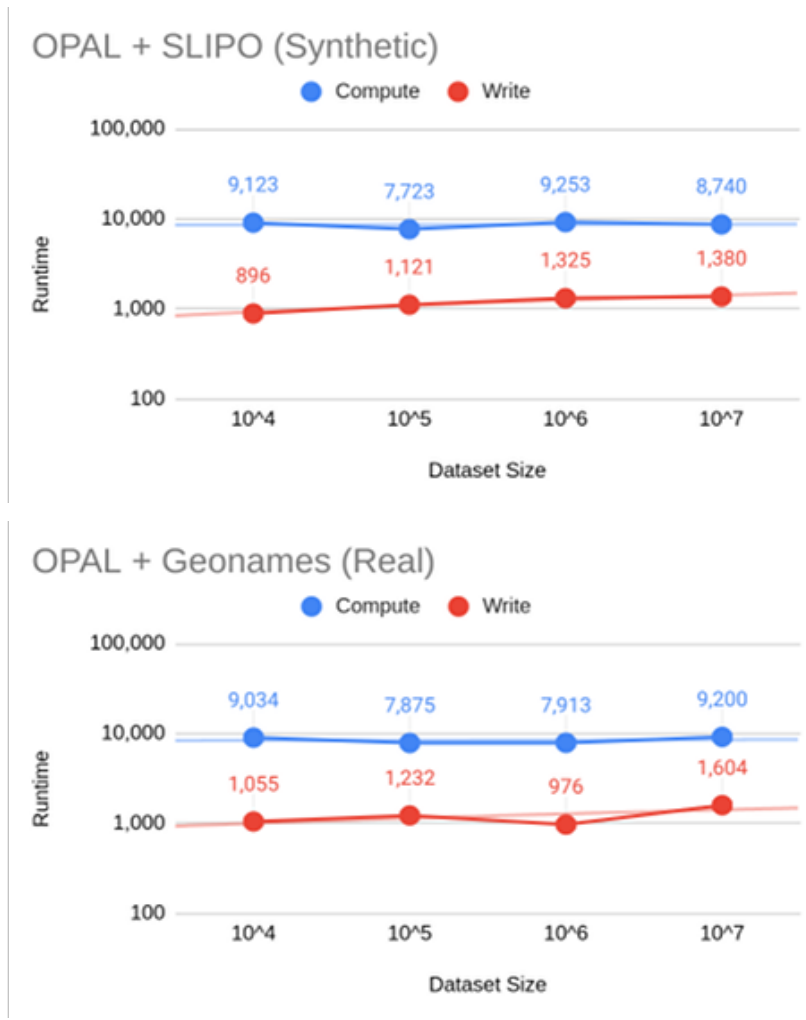

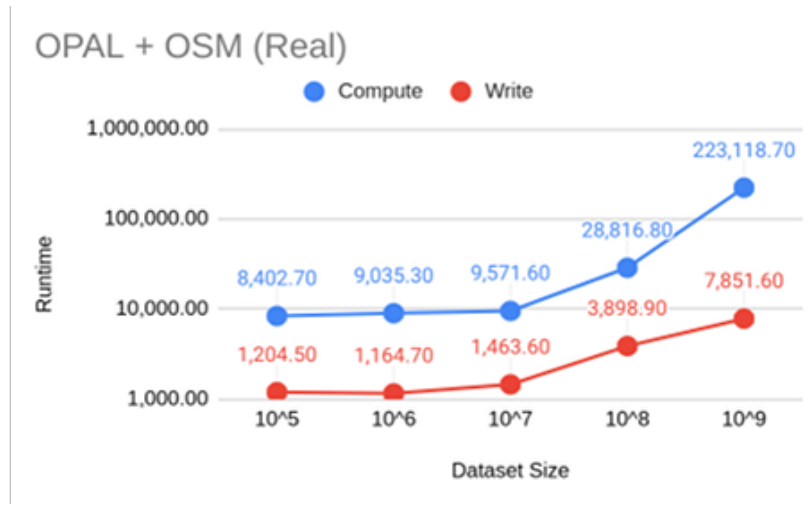

**Weiterführende Inhalte**

- Software LIMES Spark: https://github.com/dice-group/LIMES/tree/feature/hr3-spark
- ORCHID - Reduction-Ratio-Optimal Computation of Geo-Spatial Distances for Link Discovery (Artikel von Axel-Cyrille Ngonga Ngomo): https://doi.org/10.1007/978-3-642-41335-3_25
- Datensatz LinkedGeoData: http://linkedgeodata.org/

### 2.5.4 Arbeitspaket 5.4: Lizenzintegration

Um die Gefahr zu vermeiden, dass Daten mit inkompatiblen Lizenzen von Nutzern kombiniert werden, wurde in OPAL ein Ansatz entwickelt, mit dem Lizenzen über ihre Attribute auf Kompatibilität geprüft werden. Der Prototyp stellt eine Assistenzfunktion bereit, die Nutzern Informationen darüber liefert, ob mehrere Datensatz-Lizenzen kompatibel sind und welche Lizenzen für eine Relizensierung der kombinierten Datensätze verwendet werden können. Unter dem Namen ReCoDa wurde ein Prototyp erstellt und ein Artikel auf der IEEE International Conference on Semantic Computing (ICSC 2021) gemeinsam mit dem OPAL-Projekt vorgestellt. Die folgende Abbildung zeigt die Nutzeroberfläche des Prototypen:

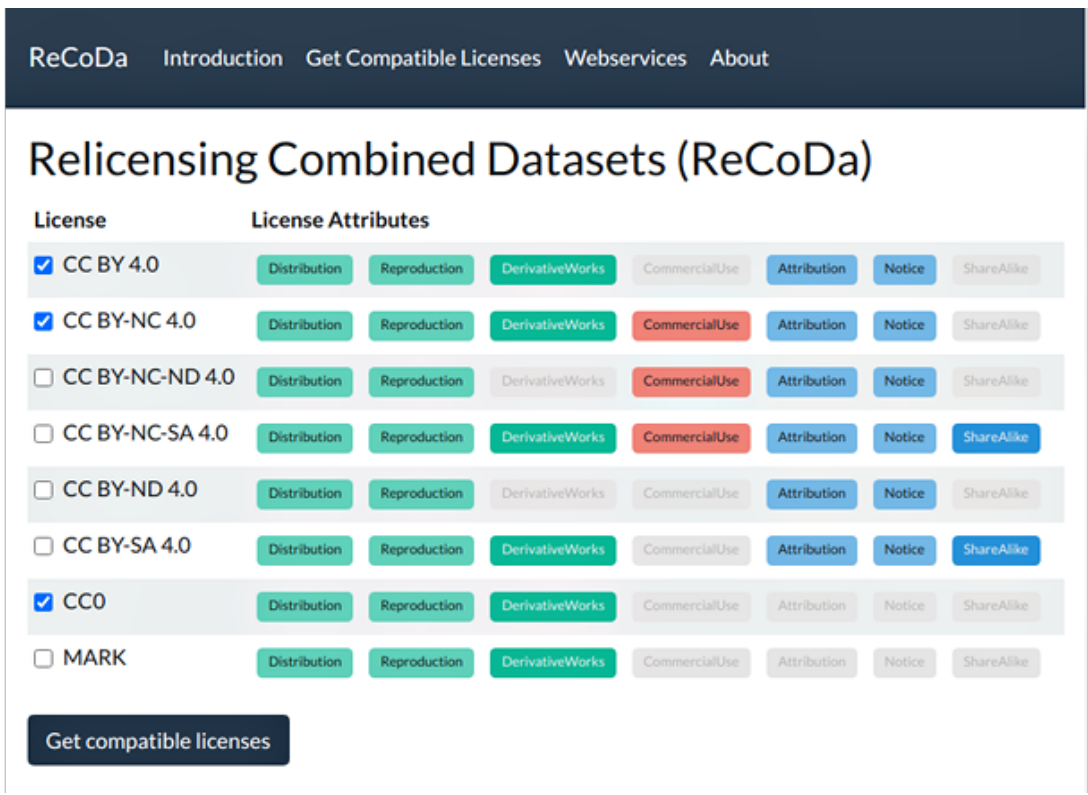

### D5.3 Lizenzen: Evaluierung

- Relicensing Combined Datasets (Artikel von Adrian Wilke, Arwa Bannoura, Axel-Cyrille Ngonga Ngomo): https://papers.dice-research.org/2021/ICSC2021_ReCoDa/Relicensing-Combined-Datasets-ReCoDa-public.pdf
- Software Demo: https://dice-research.org/ReCoDa
- Software Code: https://github.com/dice-group/ReCoDa

## 2.6 Arbeitspaket 6: Datenselektion

**Ziel des Arbeitspaketes**

Relevante Teile eines Datensatzes lassen sich anhand von Prädikaten und Relationen sowie räumlichen Relationen auswählen, um den Umfang der übertragenen Daten möglichst zu minimieren.

**Deliverables des Arbeitspaketes**

Die folgenden Abgaben wurden in diesem Arbeitspaket fertiggestellt:

- D6.1 Linked-Data-Slicing-Komponente
- D6.2 Räumliches Slicing

### 2.6.1 Arbeitspaket 6.1: Slicing von Linked Data

Zum Slicing von großen Linked Data Graphen wurden die zwei folgenden Ansätze entwickelt.

**Ansatz 1: ElasticTriples**

Dieser Ansatz verwendet die Suchmaschine Elasticsearch. Dabei werden die Tripel-Bestandteile Subjekt, Prädikat und Objekt serialisiert, separiert und in drei entsprechende Felder geschrieben. Durch iterative Anfragen können Graphen einzelner Datensätze anschließend gebildet werden.

Ein Import von 90 Millionen Triple (16.3 GB im N-Triples Format) benötigt 77 Minuten. Für das anschließende Splitting sind je Datensatzgraphgröße 2 bis 3 Sekunden notwendig. Bei einer exemplarischen Extraktion wurden 206 zugehörige Triple mit 2.281 Einzelanfragen iterativ gesplittet.

**Ansatz 2: OPAL Slicer**

Dieser Ansatz nutzt Muster im SPARQL Format um Untermengen von Wissensgraphen zu extrahieren. Für diese Lösung wurde das RDFSlice Projekt verwendet. Dabei kommt das Extract-Transform-Load (ETL) Paradigma zum Einsatz. Es werden Graphmuster verwendet, die maximal eine Variable oder einen Identifikatoren in den Join-Bedingungen des Musters verwendet. Durch diese Einschränkung entsteht ein Geschwindigkeitsvorteil gegenüber konventionellen Triplestores.

**Weiterführende Inhalte**

- Software ElasticTriples: https://github.com/projekt-opal/ElasticTriples
- Software Slicer: https://github.com/projekt-opal/slicer

### 2.6.2 Arbeitspaket 6.2: Slicing von Geodaten

Die Indexierung und Speicherung von angereicherten Geodaten geschieht aufbauend auf der Suchmaschine Elasticsearch. Zum Zugriff dieser Daten wurden OPAL Webservices eingerichtet, die die Elasticsearch Geodaten API verwenden. Nach Angabe einer BoundingBox, bestehend aus zwei Koordinaten, werden die im ausgewählten Bereich verfügbaren Datensätze zurückgegeben.

Im Rahmen des OPAL Open Data Hackathon wurde eine bestehende Softwarekomponente erweitert und an die Anforderungen des OPAL Portals angepasst. In einer Web-Komponente basierend auf OpenLayers wird eine Karte angezeigt. Portalnutzer können eine BoundingBox auf der Karte aufziehen. Datensätze, deren Koordinaten sich innerhalb der Markierung befinden, werden entsprechend im OPAL Portal aufgelistet.

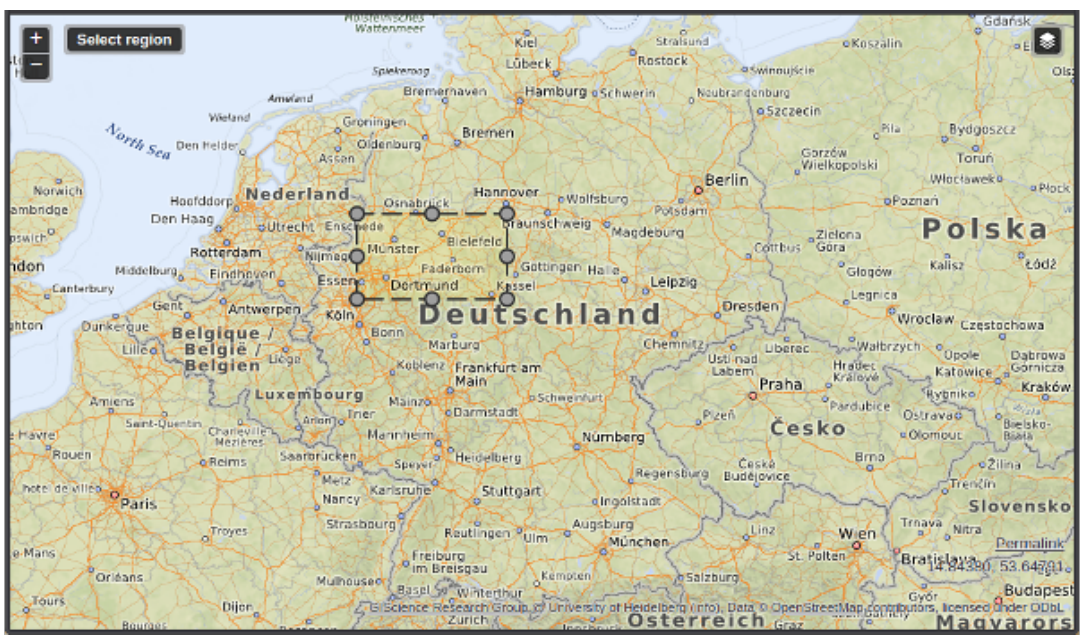

**Weiterführende Inhalte**

- Software Spatial Slicing: https://github.com/projekt-opal/hackathon/tree/gh-pages/geo
- OpenLayers: https://openlayers.org/

## 2.7 Arbeitspaket 7: Anwendungsfälle

**Ziel des Arbeitspaketes**

Die Anwendbarkeit des Linked-Data-Ansatzes zur Extraktion und Verwaltung von Metadaten offener Datensätze soll anhand der Suchfunktion als zentraler Komponente eines Datenportals sowie weiterer Demonstratoren validiert werden.

**Deliverables des Arbeitspaketes**

Die folgenden Abgaben wurden in diesem Arbeitspaket fertiggestellt:

- D7.1 Suchkomponente
- D7.2 Benchmarking der Suchkomponente
- D7.3 City-App Demonstrator
- D7.4 Social Media Bot Demonstrator

### 2.7.1 Arbeitspaket 7.1: Suche

Auch mit der Indexierung von Volltexten bleibt das Problem der Verwendung von Suchbegriffen bestehen, die nicht in den ursprünglichen Metadaten enthalten sind. Obwohl ein Anwender semantisch korrekte Begriffe verwendet, werden potenziell verfügbaren Datensätze nicht gefunden. Beispielsweise könnte ein Volltext das Wort "Stadtbahn" enthalten. Die Verwendung von Suchbegriffen wie "S-Bahn" oder "Straßenbahn" würde den entsprechenden Datensatz ggf. nicht finden. Um die Auffindbarkeit zu verbessern können Synonyme (unterschiedliche Begriffe ähnlicher Bedeutung) aufgelöst werden. Hierdurch werden zusätzliche semantisch verwandte Begriffe in die Suche einbezogen.

Zur Verbesserung der Auffindbarkeit von Datensätzen werden Begriffe einbezogen, die nicht direkter Bestandteil von Volltexten sind. Dies geschieht über eine Generierung von Synonym-Listen, die anschließend in die Suche eingebunden werden können. Die Extraktion der Synonyme geschieht über eine SPARQL Anfrage und folgender Verfeinerung:

1. Filterung aller Nomen in deutscher Sprache,
2. Einschränkung auf die deutschsprachigen Nomen, für die Synonyme spezifiziert sind,
3. Extraktion der kanonischen Form der einzelnen deutschsprachigen Nomen und
4. Extraktion der kanonischen Formen der jeweiligen Synonyme.

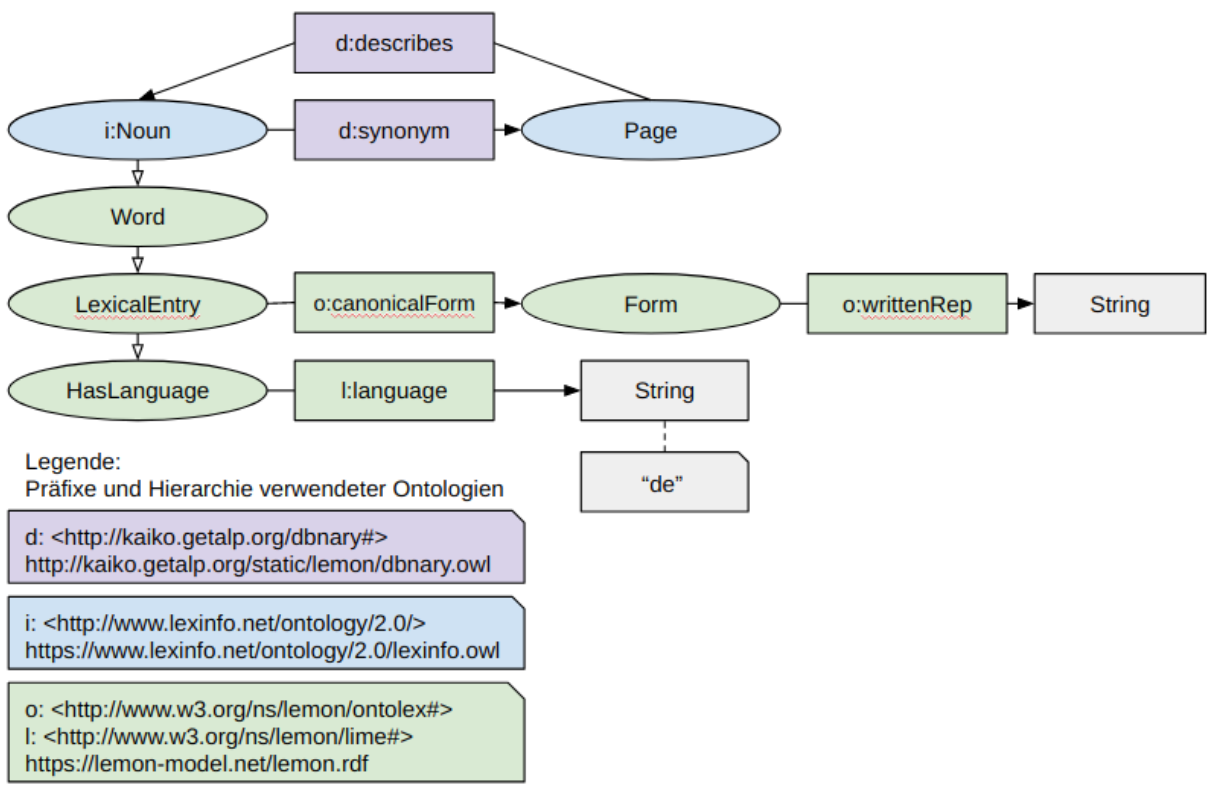

Als Ergebnis wurden 6.668 deutschsprachige Nomen extrahiert, für die Synonyme bekannt sind. Zu den entsprechenden Nomen wurden insgesamt 21.634 Synonyme zu den entsprechenden Nomen gefunden. Abschließend wurden die extrahierten Nomen mit den Titeln und Beschreibungstexten der Metadaten aus mCLOUD und GovData abgeglichen und damit die Anzahl relevanter Suchdaten eingeschränkt. Final stehen 1.497 Nomen aus mCLOUD und GovData sowie die entsprechenden Synonyme bereit.

**Benchmarking der Suchkomponente**

Der Benchmarking der Suchkomponente erfolgte über einen empirischen Vergleich der Laufzeiten der Datenhaltungslösungen Elasticsearch und dem RDF Triplestore Apache Fuseki. Hierzu wurden annähernd 60.000 Datensätze in den Varianten gespeichert. Anschließend fand eine Laufzeitanalyse über fünf Abfragen statt, die jeweils 100 mal ausgeführt wurden. Danach wurden das arithmetische Mittel sowie die Standardabweichungen errechnet. Als Ergebnis lässt sich festhalten, dass einfache Abfragen rund 4 mal performanter mit Elasticsearch durchgeführt wurden. Erweiterte Abfragen (mit SPARQL Filtern) erfolgten 5 bis 20 mal performanter mit Elasticsearch.

**Weiterführende Inhalte**

- D7.1 Suchkomponente (Adrian Wilke, Caglar Demir): https://github.com/projekt-opal/doc/blob/master/deliverables/OPAL_D7.1_Suchkomponente.pdf
- D7.2 Benchmarking der Suchkomponente (Caglar Demir, Adrian Wilke): https://github.com/projekt-opal/doc/blob/master/deliverables/OPAL_D7.2_Search-component-benchmark.pdf
- Software Search-Component-Benchmark: https://github.com/projekt-opal/Search-Component-Benchmark

### 2.7.2 Arbeitspaket 7.2: Mobile App

Für den Demonstrator der mobilen Applikation wurde zwischen nativen Lösungen für Smartphones und einer Webkomponente abgewogen. Da der Nutzen nativer Lösungen im Verhältnis zum Entwicklungsaufwand als gering eingeschätzt wurde, viel die Entscheidung zugunsten einer Webapplikation aus. Die Implementierung geschah als Integration in das OPAL Webportal. Das Portal wurde dementsprechend mit Responsivem Webdesign gestaltet, so dass eine optimierte Ansicht für mobile Endgeräte verfügbar ist. Nutzer können bei der Sortierung von Datensätzen einen Standort eingeben oder diesen über die W3C Geolocation API Spezifikation vom Gerät abfragen lassen. Anschließend werden Datensätze nach aufsteigendem Abstand zum Standort aufgelistet, so dass eine Identifikation von Daten am aktuellen Standort erfolgt und relevante Datensätze im Umfeld angezeigt werden.

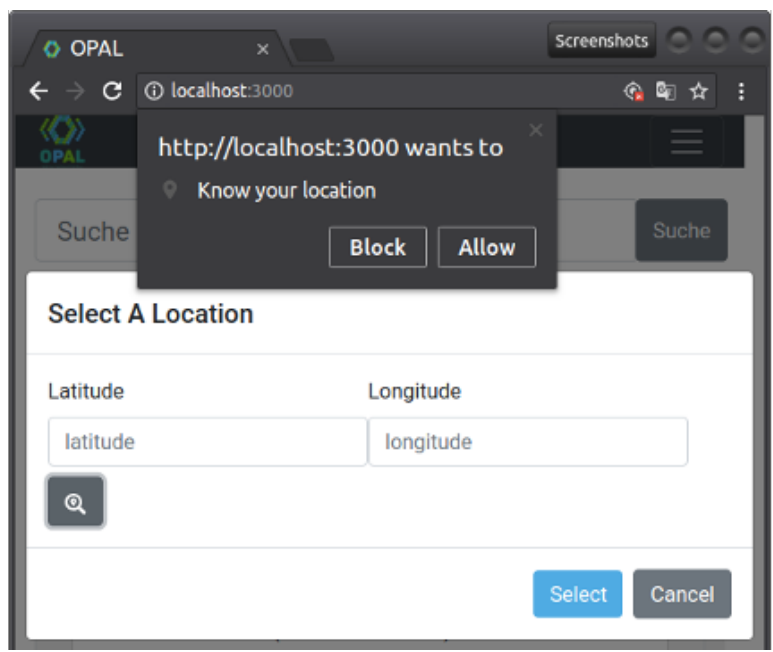

**Weiterführende Inhalte**

- Software OPAL Web UI: https://github.com/projekt-opal/web-ui
- W3C Geolocation API Spezifikation: https://w3c.github.io/geolocation-api/

### 2.7.3 Arbeitspaket 7.3: Social Bot

Die Social Bot Komponente wurde innerhalb einer Bachelorarbeit verwirklicht. Dabei wurde, im Gegensatz zu generischen Entwicklungen, ein Question Answering System entwickelt, dass speziell auf die unterliegende DCAT-Datenstruktur abzielt. Es wurde ein Template-basierter Ansatz umgesetzt, mit dem natürlichsprachige Anfragen analysiert und Entitäten erkannt werden. Anschließend wird ein passendes Template selektiert, Dantenbank-Anfragen ausgeführt und die entsprechenden Ergebnisse zurückgegeben. Als Nutzerschnittstellen wurden eine Web-Ansicht sowie eine Twitter-Anbindungen realisiert. Die Arbeit wurde von Mitarbeitern aus den Projekten OPAL und LIMBO betreut.

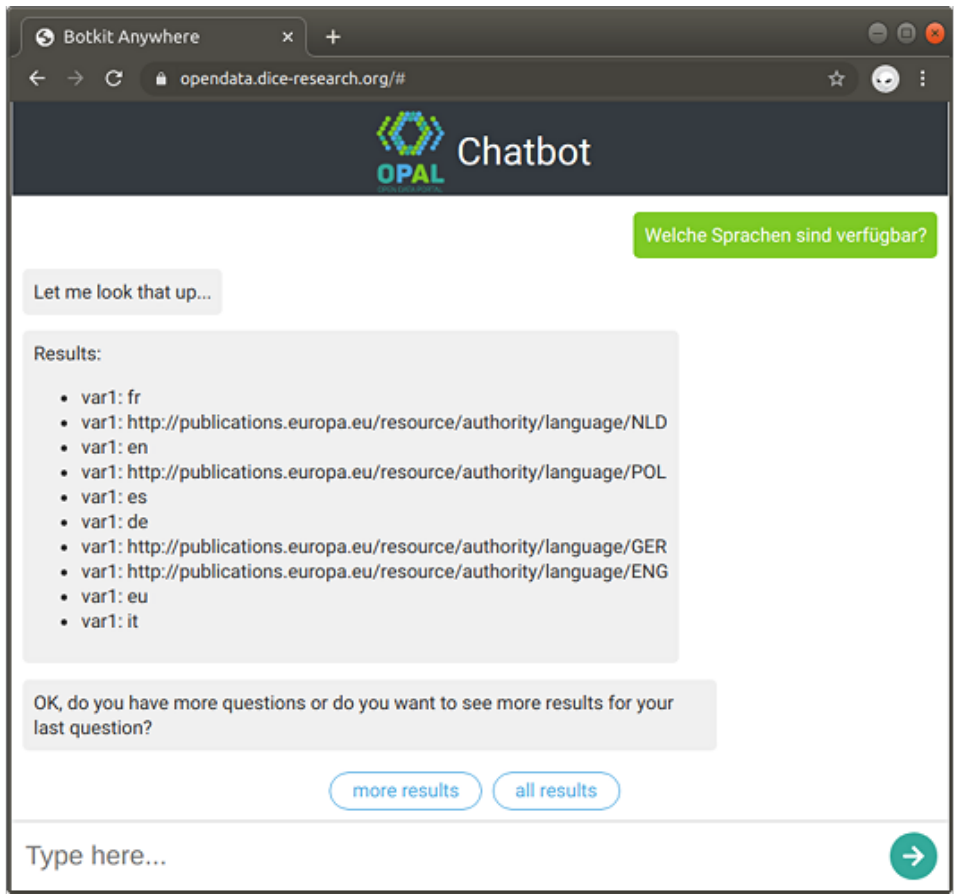

**Weiterführende Inhalte**

- D7.3 Social Media Bot Demonstrator "A Question Answering (QA) System for the Data Catalog Vocabulary (DCAT)" (Marten Louis Schmidt, Daniel Vollmers, Adrian Wilke): https://github.com/projekt-opal/dcat-qa/blob/thesis/thesis.pdf
- Software DCAT QA: https://github.com/projekt-opal/dcat-qa

## 2.8 Arbeitspaket 8: Portalentwicklung

**Ziel des Arbeitspaketes**

In Arbeitspaket 8 wird das Open Data Portal Germany als erweiterbare komponentenbasierte Webanwendung entwickelt.

**Deliverables des Arbeitspaketes**

Die folgenden Abgaben wurden in diesem Arbeitspaket fertiggestellt:

- D8.1 Portalinfrastruktur
- D8.2 Erster Portalprototyp
- D8.3 Erweiterter Portaldemonstrator
- D8.4 Finales OPAL-Portal
- D8.5 Anwenderdokumenation zum OPAL-Portal
- D8.6 Evaluierungsergebnisse

### 2.8.1 Arbeitspaket 8.1: Portalinfrastruktur und Deployment

Das Deployment der Komponenten des OPAL Portals geschieht über Docker. Die primäre Datenhaltung für das Portal wurde mit der performanten Suchmaschine Elasticsearch realisiert. Für Datenanalysen werden die Metadaten zusätzlich im Triplestore Fuseki vorgehalten. Beide Datenhaltungskomponenten verfügen über Schnittstellen (API, SPARQL) zum Datenzugriff. Diese werden über Microservices zugegriffen. Die Microservices wurden mit Webservices des Spring Frameworks implementiert und bereiten Daten für die Verwendung in der Benutzeroberfläche (Web UI) auf. Die Benutzeroberfläche baut auf den Bibliotheken React, Next.js und Bootstrap auf. Die folgende Abbildung gibt einen Überblick über die Hauptkomponenten und Repositorien:

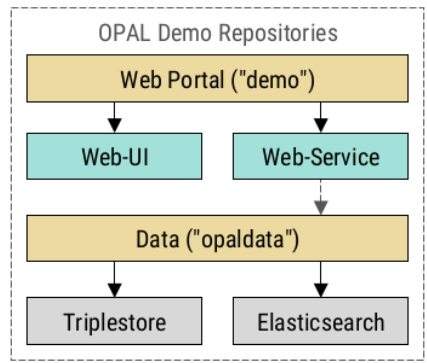

**Weiterführende Inhalte**

- 8.1 Portalinfrastruktur (Matthias Wauer, Afshin Amini, Adrian Wilke): https://github.com/projekt-opal/doc/blob/master/deliverables/OPAL_D8.1_Portal_infrastructure.pdf
- Software Datenhaltung: https://github.com/projekt-opal/opaldata
- Software Microservices: https://github.com/projekt-opal/web-service
- Software Benutzeroberfläche: https://github.com/projekt-opal/web-ui

### 2.8.2 Arbeitspaket 8.2: Integration des Portalsystems

Der erste Portalprototyp (D8.2) basiert auf der Datenkatalog-Software CKAN. Während des Projektverlaufs stellte sich heraus, dass die Geschwindigkeit des Datenimports sowie die unterliegende relationale Datenbank nicht den Projektanforderungen entsprachen. Zur Datenanalyse erfolgte eine Umstellung auf RDF. Der folgende Screenshot zeigt den ersten Protalprototyp:

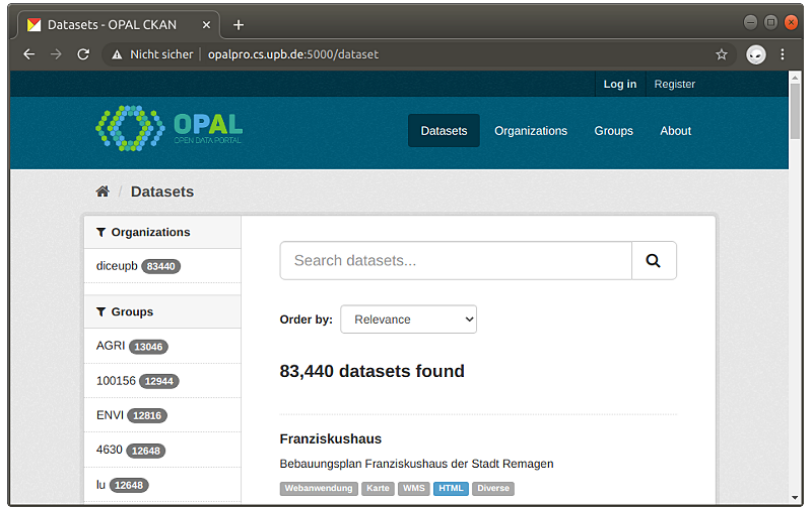

Für den erweiterten Portaldemonstrator (D8.3) wurde eine Architektur basierend auf den oben beschriebenen Technologien (Arbeitspaket 8.1) entwickelt. Die folgende Abbildung zeigt einen Entwurf der geplanten Frontend-Elemente und Filtermöglichkeiten. Diese wurden in Arbeitspaket 8.3 umgesetzt.

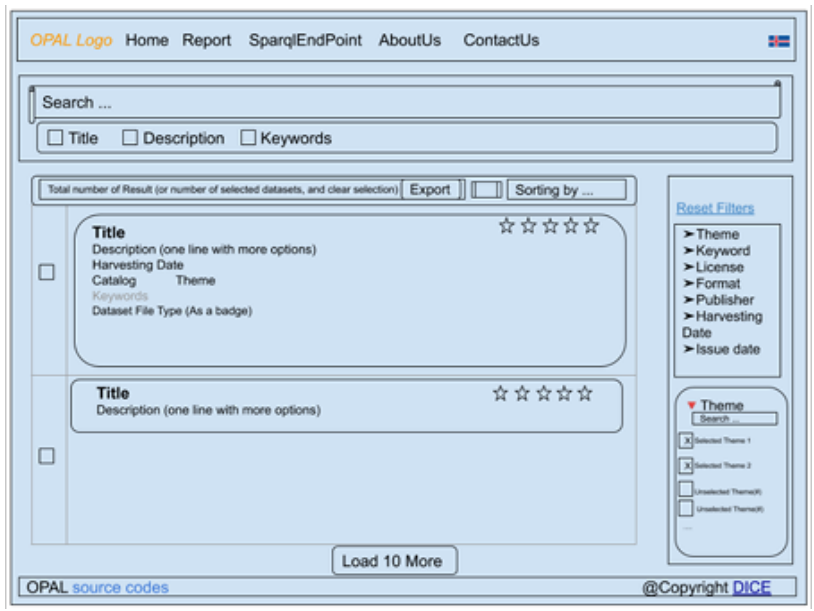

**Weiterführende Inhalte**

- D8.2 Erster Portalprototyp (Zafar Habeeb Syed, Afshin Amini, Adrian Wilke, Matthias Wauer): https://github.com/projekt-opal/doc/blob/master/deliverables/OPAL_D8.2_First_portal_prototype.pdf
- D8.3 Erweiterter Portaldemonstrator (Afshin Amini, Adrian Wilke): https://github.com/projekt-opal/doc/blob/master/deliverables/OPAL_D8.3_Enhanced_portal_demonstrator.pdf
- Software CKAN: https://github.com/projekt-opal/docker-ckan

### 2.8.3 Arbeitspaket 8.3: Webportal

**OPAL Portal**

Das finale OPAL-Portal wurde mit den Technologien aus Arbeitspaket 8.1 und dem Entwurf aus Arbeitspaket 8.2 umgesetzt. Das zentrale Element ist die Suchfunktion, mit der Titel, Beschreibungstexte und Stichwörter durchsucht werden können. Metadatensätze können nach Kategorien, Herausgebern, Katalogen und Lizenzen gefiltert werden. Zusätzlich zur Anzeige der größten Datenquellen und -herausgebern sind die mobile App sowie das räumliche Slicing implementiert.

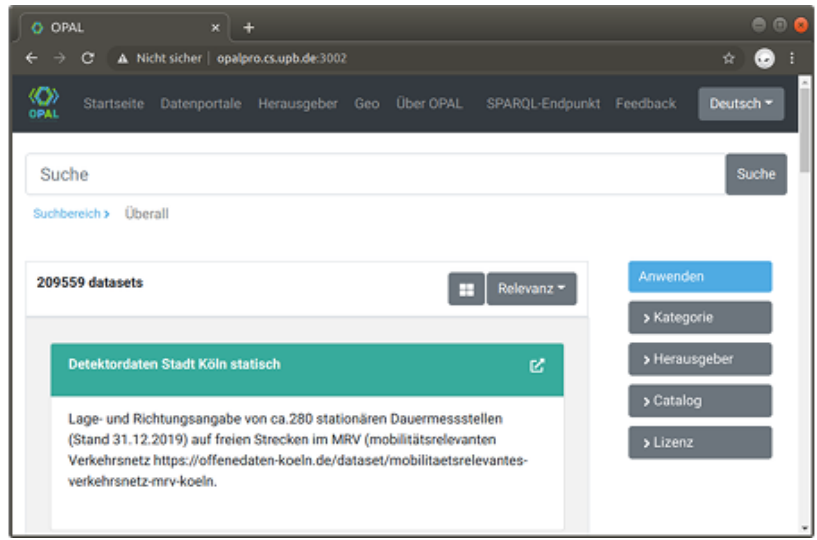

**Anwenderdokumentation**

Hinsichtlich der Weiterverwertung der Projektergebnisse ist eine Dokumentation für entsprechende Anwender notwendig. Zur mittel- und langfristigen Auffindbarkeit wurden die entsprechenden Produkte auf verschiedenen Online-Plattformen zugänglich gemacht. Die wissenschaftlichen Ergebnisse wurden als Dokumente im PDF Format veröffentlicht und sind sowohl auf der DICE Projektwebseite als auch über die URLs in diesem Dokument verfügbar.

Über die Projektwebseite sind außerdem die erstellten Deliverables aus den Arbeitspaketen, Quelldaten, verfeinerte Daten sowie das OPAL Webportal offen abrufbar.

Zur Wiederverwendung und Weiterentwicklung der Softwarekomponenten steht mit D8.5 eine technische Anwenderdokumenation bereit.

Im Projektverlauf wurden rund 50 Code-Repositorien erstellt. Zur Dokumentation fand eine Aufteilung in Haupt- und Nebenprodukte statt, die mit Kurzbeschreibungen auf der Repositorien Dokumentation aufgeführt sind.

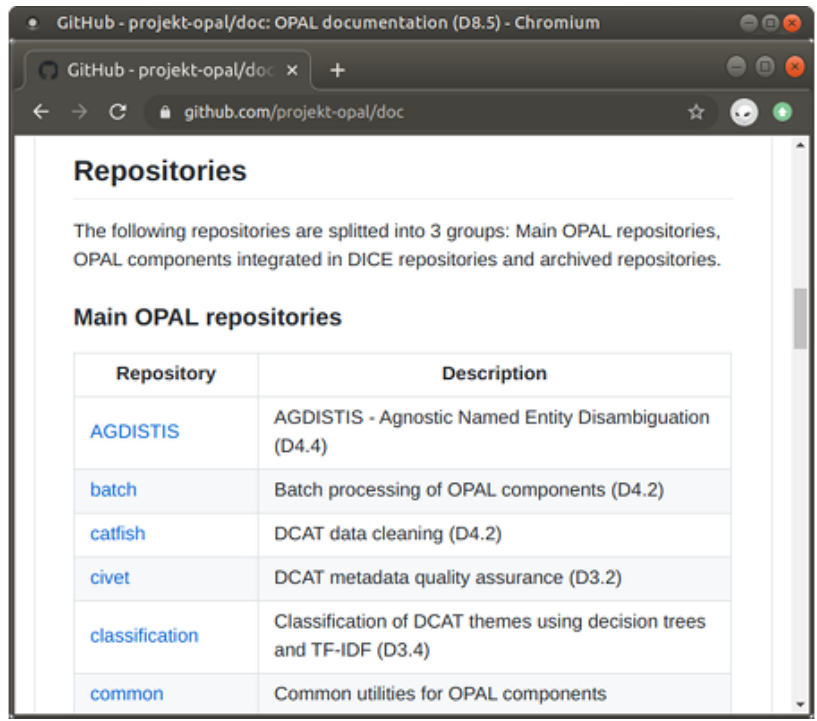

**Weiterführende Inhalte**

- Demonstrator OPAL Portal: https://dice-research.org/OPAL-Demo
- Software OPAL Portal: https://github.com/projekt-opal/demo
- D8.5: Anwenderdokumenation zum OPAL-Portal (Adrian Wilke): https://github.com/projekt-opal/doc/blob/master/deliverables/OPAL_D8.5_Documentation.pdf
- Dokumentation Repositorien: https://github.com/projekt-opal/doc#repositories
- DICE Projektwebseite: https://dice-research.org/OPAL

### 2.8.4  Arbeitspaket 8.4: Evaluation von Portal und Anwendungen

Zur Evaluation des Webportals wurde eine Umfrage basierend auf System Usability Scale (SUS) durchgeführt. SUS ist ein einfacher und technologieunabhängiger Fragebogen, mit dem eine Bewertung der Gebrauchstauglichkeit von Systemen vorgenommen wird. Er besteht aus 10 Fragen nach der Likert-Skala, die von *Stimme überhaupt nicht zu (1)* bis *Stimme voll und ganz zu (5)* reicht. Die nachfolgende Abblildung zeigt das Ergebnis der Umfrage, welches eine positive Resonanz von Anwendern darstellt:

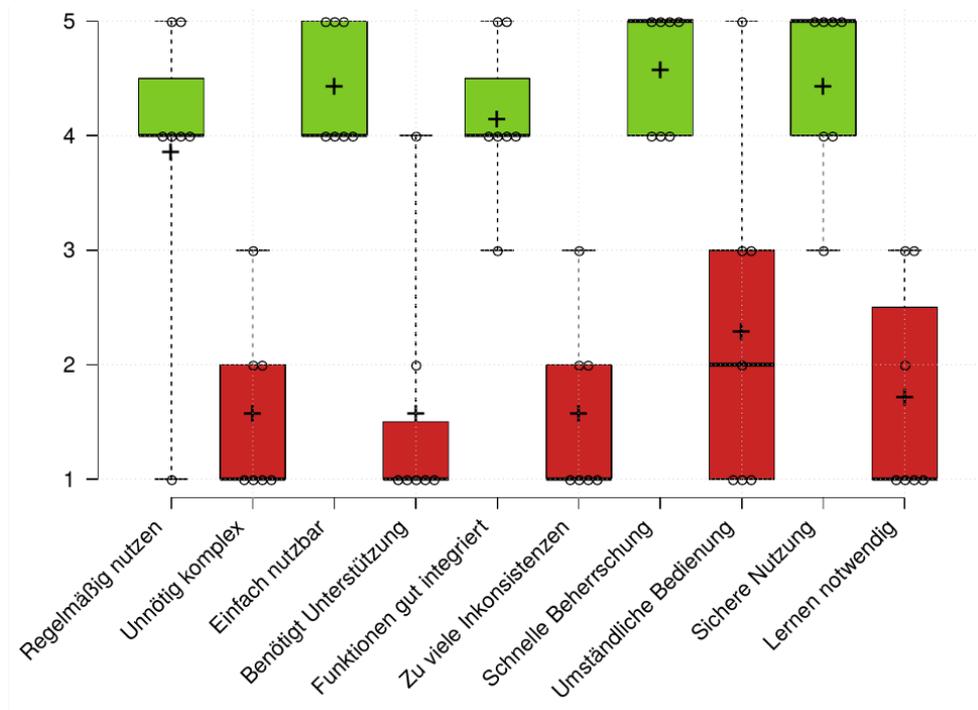

Eine technische Evaluierung fand in den Deliverables D2.3/D2.6 (Crawler Benchmark), D4.3 (Indexstrukturen) und D7.2: (Benchmark Suchkomponente) statt.

**Weiterführende Inhalte**

- D8.6 Evaluierungsergebnisse (Arwa Bannoura, Adrian Wilke): https://github.com/projekt-opal/doc/blob/master/deliverables/OPAL_D8.6_Evaluation.pdf

## 2.9 Arbeitspaket 9: Projektmanagement

**Ziel des Arbeitspaketes**

Dieses Arbeitspaket beinhaltet die Koordination, Dokumentation und Überwachung der Meilensteine des Projekts, die Kommunikation mit dem Projektträger sowie die Kommunikation mit Externen aus Industrie und Forschung.

**Deliverables des Arbeitspaketes**

Die folgenden Abgaben wurden in diesem Arbeitspaket fertiggestellt:

- D9.1 Kommunikations- und Disseminationsplan
- D9.2 Projektbericht Jahr 1
- D9.3 Projektbericht Jahr 2
- D9.4 Projektbericht Jahr 3

### 2.9.1 Arbeitspaket 9: Projektmanagement

Die Projektberichte für die entsprechenden Zeitpunkte wurden eingereicht.

**Kommunikation und Dissemination**

Ein Austausch und eine Vernetzung mit Unternehmen und Wissenschaftlern aus dem Bereich Mobilitätsdaten fand in erster Linie über Teilnahmen an mFUND Workshops statt. Zudem wurde das OPAL Projekt bereits auf der mFUND-Konferenz 2018 vorgestellt.

Die Dissemination in der Wissenschaft fand über rund 30 Publikationen statt. Auf den jeweiligen Konferenzen wurde das OPAL Projekt in Bezug auf die jeweiligen Konferenzbeiträge vorgestellt. In den zugehörigen Publikationen ist das Projekt mit Förderkennzeichen jeweils aufgeführt.

Die Zielgruppe der Studierenden wurde primär im Rahmen des OPAL Open Data Hackathon am 6. April 2020 einbezogen. Der ursprünglich geplante Hackathon musste aufgrund der Corona-Pandemie umgestellt werden und fand online statt.

Die allgemeinen Kommunikation und Dissemination fand über die Veröffentlichung von Blogbeiträgen statt. Es wurden die Themen mFUND-Konferenz und -Workshops, Datenanalysen, wissenschaftliche Konferenzbeiträge, Hackathon und Abschlusspräsentation vorgestellt. Zudem konnte die Reichweite der DICE Fachgruppe auf Twitter zur Dissemination von Projektergebnissen beitragen.

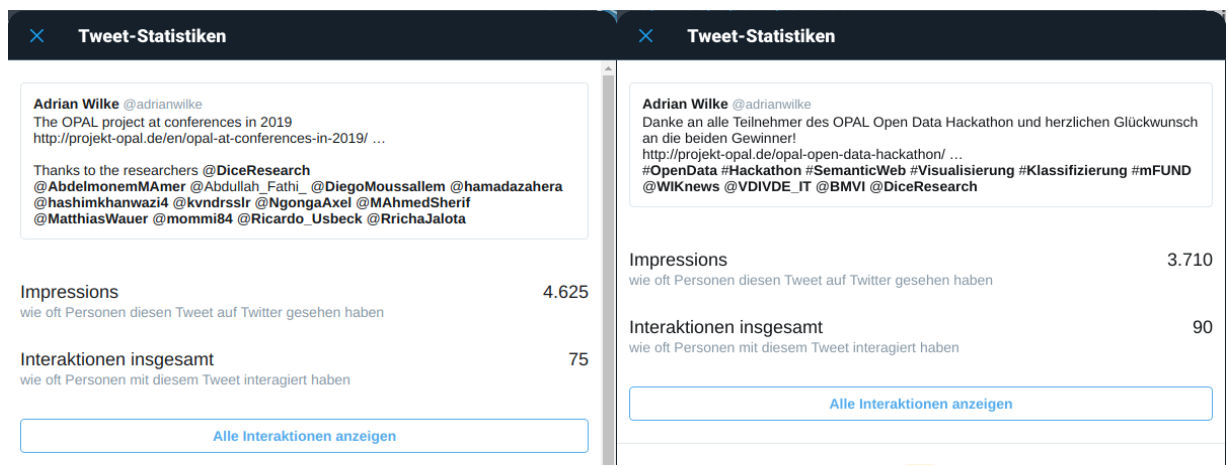

**Weiterführende Inhalte**

- D9.1: Kommunikations- und Disseminationsplan (Matthias Wauer): https://github.com/projekt-opal/doc/blob/master/deliverables/OPAL_D9.1_Communication_and_dissemination.pdf
- OPAL Open Data Hackathon: https://projekt-opal.github.io/hackathon/
- Neuigkeiten Blog: http://projekt-opal.de/category/allgemein/
- Twitter Suche OPAL: https://twitter.com/search?q=OPAL%20%20(%40DiceResearch)

# 3 Projektstruktur: Meilensteine, Arbeitspakete und Deliverables

Für das Projekt sind die folgenden **4 Meilensteine** definiert:

- **Meilenstein 1** [M1]: **Anforderungserfassung und Architektur** (Monat 6)
- **Meilenstein 2** [M2]: **Erster Portalprototyp** (Monat 18)
- **Meilenstein 3** [M3]: **Erweiterter Portaldemonstrator** (Monat 24)
- **Meilenstein 4** [M4]: **Finales OPAL-Portal** (Monat 36)

Das Gesamtprojekt umfasst **9 Arbeitspakete** und weitere Unterteilungen. Zu diesen gibt es insgesamt **39 Deliverables (Abgaben)**. Diese sind bedingt durch inhaltliche Überschneidungen teilweise mehreren Arbeitspaketen zugeordnet. Die Strukturierung der Abeitspakete, Meilensteine und Deliverables ist im Folgenden aufgelistet:

- **Arbeitspaket 1: Anforderungsanalyse und Architektur** [M2]

  - **Arbeitspaket 1.1: Erhebung der Nutzeranforderungen** [M1]
    * Deliverable 1.1: Anforderungsanalyse
  - **Arbeitspaket 1.2: Datenanalyse** [M1]
    * Deliverable 1.2: Datenanalyse
  - **Arbeitspaket 1.3: Architektur** [M2]
    * Deliverable 1.3: Architektur

- **Arbeitspaket 2: Datenakquisition** [M4]

  - **Arbeitspaket 2.1: Crawler-Spezifikation** [M2]
    * Deliverable 2.1: Spezifikation der Crawler-Komponente
  - **Arbeitspaket 2.2: Crawling-Komponente** [M4]
    * Deliverable 2.2: Erste Version der Crawler-Komponente
    * Deliverable 2.5: Finale Crawler-Komponente
  - **Arbeitspaket 2.3: Metadatenbasierte Crawlingstrategien** [M4]
    * Deliverable 2.4: Metadatenbasierte Crawlingstrategien
  - **Arbeitspaket 2.4: Benchmarking der Crawling-Komponente** [M4]
    * Deliverable 2.3: Benchmark-Spezifikation und Ergebnisse des ersten Crawlers
    * Deliverable 2.6: Finale Crawler-Benchmark-Ergebnisse

- **Arbeitspaket 3: Datenanalyse** [M4]

  - **Arbeitspaket 3.1: Qualitätsanalyse** [M4]
    * Deliverable 3.1: Spezifikation von Qualitätskriterien
    * Deliverable 3.2: Qualitätsanalyse-Komponente
    * Deliverable 3.5: Finale Datenanalysekomponenten
  - **Arbeitspaket 3.2: Semistrukturierte Metadatenextraktion** [M4]

* Deliverable 3.3: Erste Metadatenextraktionskomponente
* Deliverable 3.5: Finale Datenanalysekomponenten

– **Arbeitspaket 3.3: Unstrukturierte Metadatenextraktion** [M4]

* Deliverable 3.3: Erste Metadatenextraktionskomponente
* Deliverable 3.5: Finale Datenanalysekomponenten

– **Arbeitspaket 3.4: Topic- und Schema-Extraktion** [M4]

* Deliverable 3.4: Topic-Extraktionskomponente
* Deliverable 3.5: Finale Datenanalysekomponenten

• **Arbeitspaket 4: Datenkonvertierung** [M4]

– **Arbeitspaket 4.1: Vokabularspezifikation** [M2]

* Deliverable 4.1: Vokabularspezifikation

– **Arbeitspaket 4.2: Konvertierungskomponente** [M4]

* Deliverable 4.2: Konvertierungskomponente

– **Arbeitspaket 4.3: Indexstrukturen** [M3]

* Deliverable 4.3: Prototyp Indexstrukturen und Entitätserkennung

– **Arbeitspaket 4.4: Indizierungskomponente** [M4]

* Deliverable 4.4: Indizierungskomponente

• **Arbeitspaket 5: Datenintegration** [M4]

– **Arbeitspaket 5.1: Datenverknüpfung** [M4]

* Deliverable 5.1: Erste Version der Verknüpfungskomponente
* Deliverable 5.5: Finale Verknüpfungskomponente

– **Arbeitspaket 5.2: Lernalgorithmen für Linkspezifikationen** [M4]

* Deliverable 5.2: Lernalgorithmen für Linkspezifikationen auf Metadaten

– **Arbeitspaket 5.3: Erweiterte Lernalgorithmen über Daten** [M4]

* Deliverable 5.4: Erweiterte Lernalgorithmen für Linkspezifikationen auf Metadaten

– **Arbeitspaket 5.4: Lizenzintegration** [M4]

* Deliverable 5.3: Lizenzintegrationskomponente

• **Arbeitspaket 6: Datenselektion** [M4]

– **Arbeitspaket 6.1: Slicing von Linked Data** [M4]

* Deliverable 6.1: Linked-Data-Slicing-Komponente

– **Arbeitspaket 6.2: Slicing von Geodaten** [M4]

* Deliverable 6.2: Räumliches Slicing

• **Arbeitspaket 7: Anwendungsfälle** [M4]

– **Arbeitspaket 7.1: Suche** [M4]

* Deliverable 7.1: Suchkomponente
* Deliverable 7.2: Benchmarking der Suchkomponente

– **Arbeitspaket 7.2: Mobile App** [M4]

* Deliverable 7.3: City-App Demonstrator

– **Arbeitspaket 7.3: Social Bot** [M4]

* Deliverable 7.4: Social Media Bot Demonstrator

• **Arbeitspaket 8: Portalentwicklung** [M4]

– **Arbeitspaket 8.1: Portalinfrastruktur und Deployment** [M4]

* Deliverable 8.1: Portalinfrastruktur
* Deliverable 8.4: Finales OPAL-Portal

– **Arbeitspaket 8.2: Integration des Portalsystems** [M4]

* Deliverable 8.2: Erster Portalprototyp
* Deliverable 8.3: Erweiterter Portaldemonstrator
* Deliverable 8.4: Finales OPAL-Portal

– **Arbeitspaket 8.3: Webportal** [M4]

* Deliverable 8.4: Finales OPAL-Portal
* Deliverable 8.5: Anwenderdokumenation zum OPAL-Portal

– **Arbeitspaket 8.4: Evaluation von Portal und Anwendungen** [M4]

* Deliverable 8.6: Evaluierungsergebnisse

• **Arbeitspaket 9: Projektmanagement** [M4]

– Arbeitspaket 9: Projektmanagement

* Deliverable 9.1: Kommunikations- und Disseminationsplan
* Deliverable 9.2: Projektbericht Jahr 1
* Deliverable 9.3: Projektbericht Jahr 2
* Deliverable 9.4: Projektbericht Jahr 3

# 4 Wissenschaftliche Veröffentlichungen

- Abdullah Fathi Ahmed, Mohamed Ahmed Sherif, Diego Moussallem, Axel-Cyrille Ngonga Ngomo.
  **Multilingual verbalization and summarization for explainable link discovery**.
  Data & Knowledge Engineering, page 101874, 2021. https://papers.dice-research.org/2021/DATA
  K_LSVS_journal/public.pdf.
- Abdullah Fathi Ahmed, Mohamed Ahmed Sherif, Axel-Cyrille Ngonga Ngomo.
  **Do your resources sound similar? on the impact of using phonetic similarity in link discovery**.
  In K-CAP 2019: Knowledge Capture Conference, 2019. https://papers.dice-research.org/2019/KC
  AP_PHONETICSIMILARITY/public.pdf.
- Abdullah Fathi Ahmed, Mohamed Ahmed Sherif, Axel-Cyrille Ngonga Ngomo.
  **LSVS: Link specification verbalization and summarization. In 24th International Conference on Applications of Natural Language to Information Systems (NLDB 2019)**.
  Springer, 2019. https://papers.dice-research.org/2019/NLDB_LSVS/paper/public.pdf.
- Ram G. Athreya, Axel-Cyrille Ngonga Ngomo, Ricardo Usbeck.
  **Enhancing community interactions with data-driven chatbots-the dbpedia chatbot**.
  In Companion of the The Web Conference 2018 on The Web Conference 2018, WWW 2018, Lyon , France, April 23-27, 2018, page 143–146, 2018. https://svn.aksw.org/papers/2018/WWW_dbpedia _chatbot/public.pdf.
- Caglar Demir, Axel-Cyrille Ngonga Ngomo.
  **A physical embedding model for knowledge graphs**.
  In Xin Wang, Francesca Alessandra Lisi, Guohui Xiao, Elena Botoeva, editors, Semantic Technology, pages 192–209, Cham, 2020. Springer International Publishing. https://papers.dice-research.org /2019/JIST2019_PYKE/public.pdf.
- Abdelmoneim Amer Desouki, Felix Conrads, Michael Röder, Axel-Cyrille Ngonga Ngomo.
  **Synthg: Mimicking rdf graphs using tensor factorization**.
  In Proceedings of the 15th IEEE International Conference on Semantic Computing (ICSC), pages 76–79. IEEE Computer Society, 2021. https://papers.dice-research.org/2021/ICSC2021_SynthG/IC SC_SynthG_public.pdf.
- Abdelmoneim Amer Desouki, Michael Röder, Axel-Cyrille Ngonga Ngomo.
  **Ranking on very large knowledge graphs**.
  In Proceedings of the 30th ACM Conference on Hypertext and Social Media, pages 163–171. ACM, 2019. https://papers.dice-research.org/2019/HT_DHARE/dhare_public.pdf.
- Kleanthi Georgala, Michael Röder, Mohamed Ahmed Sherif, Axel-Cyrille Ngonga Ngomo.
  **Applying edge-counting semantic similarities to Link Discovery: Scalability and Accuracy**.
  In Proceedings of Ontology Matching Workshop 2020, 2020. https://papers.dice-research.org/2 020/OM_hECATE/public.pdf.
- Rricha Jalota, Priyansh Trivedi, Gaurav Maheshwari, Axel-Cyrille Ngonga Ngomo, Ricardo Usbeck.
  **An Approach for Ex-Post-Facto Analysis of Knowledge Graph-Driven Chatbots – the DBpedia Chatbot**.